\documentclass[useAMS,usenatbib]{mn2e}

\usepackage{graphicx}
\bibpunct{(}{)}{;}{a}{}{,}
\usepackage{txfonts}

\newcommand{\ion}[2]{#1~{\sc #2}}

\title[FORS\,2 spectropolarimetry of Wolf-Rayet stars]{Searching for a magnetic field in Wolf-Rayet stars using FORS\,2 
spectropolarimetry\thanks{Based on
observations obtained at the European Southern Observatory,
Paranal, Chile (ESO programme Nos.~086.D-0206(A) and 088.D-0284(A)).
}}

\author[Hubrig et al.]{
S.~Hubrig$^{1}$\thanks{E-mail: shubrig@aip.de},
K.~Scholz$^{2}$,
W.-R.~Hamann$^{2}$,
M.~Sch\"oller$^{3}$,
R.~Ignace$^{4}$,
I.~Ilyin$^{1}$,
\and
K.~G.~Gayley$^{5}$,
L.~M.~Oskinova$^{2}$ \\ \\
$^{1}$ Leibniz-Institut f\"ur Astrophysik Potsdam (AIP), An der Sternwarte 16, 14482 Potsdam, Germany \\
$^{2}$ Institut f\"ur Physik und Astronomie, Universit\"at Potsdam, Karl-Liebknecht-Str.~24/25, 14476 Potsdam, Germany \\
$^{3}$ European Southern Observatory, Karl-Schwarzschild-Str.~2, 85748 Garching, Germany \\
$^{4}$ Department of Physics and Astronomy, East Tennessee State University, Johnson City, TN 37663, USA \\
$^{5}$ Department of Physics and Astronomy, University of Iowa, Iowa City, IA 52242, USA
}

\begin{document}

\date{Accepted Received; in original form}

\pagerange{\pageref{firstpage}--\pageref{lastpage}} \pubyear{2016}

\maketitle

\label{firstpage}

\begin{abstract}
To investigate if magnetic fields are present in Wolf-Rayet stars, we selected a few stars in the Galaxy 
and one in the Large Magellanic Cloud (LMC). 
We acquired low-resolution spectropolarimetric 
observations with the ESO FORS\,2 instrument during two different 
observing runs. During the first run in visitor mode, we observed  the LMC  Wolf-Rayet star BAT99\,7 and 
the stars WR\,6, WR\,7, WR\,18, and WR\,23 in our Galaxy.
The second run in service mode was focused on monitoring the star WR\,6. 
Linear polarization was recorded immediately after the observations of circular polarization.  
During our visitor observing run, the magnetic field for the cyclically variable star WR\,6 was measured at a 
significance level of 3.3$\sigma$ 
($\left<B_{\rm z} \right> = 258\pm78$\,G). 
Among the other targets, the highest value for the longitudinal magnetic field, 
$\left<B_{\rm z} \right> = 327\pm141$\,G, was measured in the LMC star BAT99\,7. 
Spectropolarimetric monitoring of the star WR\,6 revealed a sinusoidal nature of 
the $\left<B_{\rm z}\right>$ variations with the known rotation period of 3.77\,d, 
significantly adding to the confidence in the detection.
The presence of the rotation-modulated magnetic variability is also indicated in our frequency periodogram. 
The reported field magnitude suffers from significant systematic uncertainties at the factor 2 level, 
in addition to the quoted statistical uncertainties, owing to the theoretical
approach used to characterize it.
Linear polarization measurements showed no line effect in the stars, apart from WR\,6. 
BAT99\,7, WR\,7, and WR\,23 do not show variability of the linear polarization over two nights. 
\end{abstract}



\section{Introduction}
\label{sect:intr}

Magnetic fields are now believed to play an important role in the evolution of massive stars.
A Tayler-Spruit dynamo mechanism \citep{spruit2002} has been predicted to be efficient in 
radiative layers of the 
stellar interior. However, these fields perhaps do not reach the photosphere with a measurable strength.
\citet{cant2009} showed that subsurface convection zones
may be present in massive stars.
If a dynamo operates in such a zone,
a magnetic field could emerge to produce observable phenomena such as 
magnetic spots or hot plasma close to the stellar surface (e.g., \citealt{waldron2007}; \citealt{ram2014}).
\citet{maeder2005} examined the effect of magnetic fields on the transport of angular momentum
and chemical mixing, and found that the potential influence on the evolution of massive stars is drastic. Thus
magnetic fields might change our whole picture about the evolution from O stars via Wolf-Rayet (WR) stars to
supernovae/gamma-ray bursts. Hitherto, neglecting magnetic fields might be one of the reasons why models
and observations of massive-star populations are still in conflict \citep{hamann2006}.
Another potential importance of magnetic fields in massive stars concerns the dynamics of the stellar winds.
Wherever $B^2 R^2_* > \dot{M} \varv_{\infty}$,
it ushers a transition to magnetic control of the wind.
If the field is weaker than this limit,
the field lines are carried along with the wind of mass-loss rate $\dot{M}$
and (terminal) velocity $\varv_{\infty}$,
and become more or less radial (e.g.\ \citealt{udDoulaOwocki2002}).

Our long-standing interest is to understand WR stars in all their
aspects, including their magnetic fields. Classical WR stars are massive stars that have lost their outer 
layers in a relatively short time. Thus, if a
massive star has an interior dynamo as predicted by the Tayler-Spruit mechanism, the magnetized layers can 
now be displayed in the atmosphere with a field strength of many kG \citep{mullan2005} or being
even further condensed by the progressing stellar contraction.

There is indirect evidence, e.g.\ from spectral variability and X-ray emission, that magnetic fields are
present in WR atmospheres (e.g.\ \citealt{mich2014}). Also the theoretical work of 
\citet{gayley2010} predicts a degree of 
circular polarization of a few times 10$^{-4}$ for magnetic fields
of about 100\,G. The detection of magnetic fields is however difficult,
chiefly because the line spectrum is formed in the strong stellar wind.
This does not only imply a dilution of the field at the place of line
formation. The big problem is the wind broadening of the emission lines
by Doppler shifts with wind velocities of a few thousand km\,s$^{-1}$. 
In high-resolution spectropolarimetric observations, some broad spectral lines extend 
over adjacent orders, so that it is necessary to adopt the order 
shapes to get the best continuum normalization.
In view of such immense line 
broadening in WR stars, to search for a weak magnetic field, we decided to use the low-resolution 
VLT instrument FORS\,2 (FOcal Reducer low dispersion Spectrograph) mounted at the 8-m Antu 
telescope, which appears to be one of the most suitable instruments in the world offering low resolution 
and the required spectropolarimetric sensitivity. 

Our sample of WR stars selected for the search of the presence of magnetic fields includes two rather bright 
WN4 stars, WR\,6 and WR\,18, and the WC6 star WR\,23.
Since the rotation period of  WR\,6 ($P_{\rm rot}=3.77$\,d) was already determined and confirmed in the past
by several teams (e.g., \citealt{lamo1986}),
one of our goals is to assess the presence of a magnetic field and its potential
cyclical modulation over the rotation cycle, as expected for tilted dipole or split monopole geometries.
Further, from our previous comprehensive analyses using the Potsdam Wolf-Rayet model atmospheres,
we have identified a few Wolf-Rayet stars in the Galaxy and the Large Magellanic Cloud (LMC) that are in an extremely advanced
stage of their evolution, being very compact ($R<\sim2\,R_\odot$) and hot ($T_{\rm eff} > 100$\,kK). 
Having lost their outer layers,
they now expose their cores and are promising targets to search for the presence of a magnetic field. 
The sample of Galactic WR stars with such very small stellar radii is very limited (cf.\ \citealt{hamann2006}).
Among them, WR\,7 is the best candidate to be observed from the VLT observatory located on Cerro Paranal. 

Additionally, probably for reasons of stellar evolution
in the lower metallicity environment, the WN population in the LMC shows a higher 
fraction of very hot and compact WN
stars (e.g.,\ \citealt{hainich2014}). We have chosen the brightest of these LMC objects, BAT99\,7, as another target. 
BAT99\,7 has an additional peculiarity that makes it an especially promising candidate for showing a strong
magnetic field, which concerns the shape of its line profiles. These profile shapes are very round, unlike the
Gaussian shape that is usually found in WR-type emission lines. We have found such type of line profiles in only
two Galactic WR stars, but in five WN stars from the LMC. The reason for this profile shape is not clear. 
Model atmospheres are not able to
reproduce it, unless one assumes broadening by extremely fast rotation. In the case of BAT99\,7,
we estimate a $v\,\sin\,i$ of 1900\,km\,s$^{-1}$ to account for the profile shape when applying just 
flux convolution. Given
the compactness of the star, such fast rotation is not impossible but close to the breakup limit, making these
stars good GRB candidates. Rotational broadening may be enhanced by wind corotation enforced by a strong
magnetic field, while the rotation itself can strengthen the dynamo.

WR\,7 does not show such round line profiles, but is one of the few Wolf-Rayet stars that have been found to 
be X-ray active. 
The origin of X-rays from 
single-star winds is still under
debate, and can be attributed either to wind shocks from the deshadowing instability 
\citep{GayleyOwocki1995,feldmeier1997}, or Corotating Interaction Regions (CIRs) \citep{mullan1984}, or to
magnetic effects (e.g., \citealt{waldron2009}; \citealt{oskinova2009}). BAT99\,7
has not been seen in X-rays yet, but due to its large distance there is no tight upper limit either.

We note that a detailed quantitative spectral analysis of all stars in our sample
was already carried out in previous studies (e.g., \citealt{hamann2006}; \citealt{sander2012};
\citealt{shenar2014}).
Any binary suspects are excluded from our target selection to avoid confusion between the magnetic field 
measurements and effects from colliding stellar winds.

In Section~\ref{sect:data}, we give an overview of our spectropolarimetric observations and the data reduction,
followed  by the presentation of the results of the magnetic field measurements for the 
individual stars in Section~\ref{sect:meas}.
Section~\ref{sect:syn} describes our measurement error validation using a synthetic spectrum and 
Section~\ref{sect:lin} is devoted to observations of the linear polarization.
Finally, we summarize the results of our observations in Section~\ref{sect:disc}.

\section{Observations and data reduction}
\label{sect:data}

\begin{table}
\caption[]{Overview of the WR stars observed in the first run.
}
\centering
\begin{tabular}{llrc}
\hline
\hline 
\noalign{\vspace{1mm}}
\multicolumn{1}{c}{Object} &
\multicolumn{1}{c}{Type} &
\multicolumn{1}{c}{$m_{\rm V}$} &      
\multicolumn{1}{c}{Date} \\      
\noalign{\vspace{1mm}}
\hline
\noalign{\vspace{1mm}}
BAT99\,7     & WN4b            & 13.6 & 2010-12-23 and 24 \\
WR\,6        & WN4             &  6.9 & 2010-12-24 \\
WR\,7        & WN4             & 11.7 & 2010-12-23 and 24 \\
WR\,18       & WN4             & 10.6 & 2010-12-24 \\
WR\,23       & WC6             & 9.0  & 2010-12-23 and 24 \\
\hline
\label{tab:2010}
\end{tabular}
\end{table}

The spectropolarimetric data presented in this paper were obtained with the FORS\,2 instrument
\citep{Appenzeller1998} during two different 
observing runs. FORS\,2 is a multi-mode instrument equipped with
polarization analyzing optics, comprising super-achromatic half-wave and quarter-wave 
phase retarder plates, and a Wollaston prism with a beam divergence of 22$\arcsec$ in 
standard resolution mode.
The first run took place in visitor mode during the two nights in 2010 December 23 and 24.
During this run, the WR stars BAT99\,7, WR\,7, and WR\,23 were observed twice and the stars WR\,6 and WR\,18 once.
An overview of the observed targets during this visitor run including their types and visual magnitudes is presented 
in Table~\ref{tab:2010}.

\begin{table}
\caption[]{Logbook of the FORS\,2 circular polarization observations of WR\,6, including 
the modified Julian date of mid-exposure followed by the rotation phase, the number of elapsed 
rotation cycles, and the achieved signal-to-noise ratio.
}
\centering
\begin{tabular}{lllrr}
\hline
\hline 
\noalign{\vspace{1mm}}
\multicolumn{1}{c}{Date} &
\multicolumn{1}{c}{MJD} &
\multicolumn{1}{c}{$p$} &
\multicolumn{1}{c}{Rotation} &
\multicolumn{1}{c}{SNR} \\
 &
 &
 &
\multicolumn{1}{c}{cycle} &
 \\
\noalign{\vspace{1mm}}
\hline
\noalign{\vspace{1mm}}
2010 Dec.~24    & 55555.1577  & 0.560 & $-$77 & 1353 \\  
2011 Oct.~10    & 55845.2716  & 0.596 & 0    & 1037 \\
2011 Oct.~11    & 55846.2584  & 0.858 & 0    & 357 \\
2011 Nov.~14    & 55880.1638  & 0.861 & 9    & 1139 \\
2011 Dec.~7     & 55903.3172  & 0.009 & 15   & 1004 \\
2011 Dec.~10    & 55906.1257  & 0.754 & 16   & 925 \\
2011 Dec.~13    & 55909.2102  & 0.573 & 16   & 983 \\
2012 Jan.~1     & 55928.0760  & 0.583 & 21   & 686 \\
2012 Jan.~2     & 55929.0803  & 0.849 & 22   & 914 \\
2012 Jan.~5     & 55932.0686  & 0.643 & 23   & 1141 \\
2012 Jan.~6     & 55933.0511  & 0.904 & 23   & 1004 \\
2012 Jan.~7     & 55934.0611  & 0.172 & 23   & 1235 \\
2012 Jan.~8     & 55935.2491  & 0.487 & 23   & 1417 \\
\hline         
\label{tab:phase}
\end{tabular}  
\end{table} 

An examination of the data obtained during the first run revealed the strongest evidence for the presence
of a magnetic field in the 
cyclically variable star WR\,6, for which a photometric periodicity of 3.77\,d was detected by \citet{lamo1986}.
Motivated by this detection, we applied for a second observing run in service mode,
consisting of twelve randomly distributed individual observations to sample the stellar rotation period.
The detection of rotational modulation of the longitudinal magnetic field would constrain the global field geometry
necessary to support physical modeling of the spectroscopic and light variations.
The observing log of the service mode observations is presented in 
Table~\ref{tab:phase}, together with the rotation phases, the elapsed rotation cycles and the signal-to-noise ratio (SNR)
in the continuum. The rotation cycle 0 was assumed to begin at the start of our monitoring of WR\,6
in service mode, i.e.\ on 2011 October 10. As the spectra of WR stars are dominated by wind emission lines, with the 
strongest emission appearing in the line \ion{He}{ii} $\lambda$4686, special care was taken not to saturate 
this line with the achieved SNR of a few thousand per pixel.
During both runs, for all targets, linear polarization was recorded immediately after the observations of
circular polarization. 

For all spectropolarimetric observations, we used the grism 600B, which has an average spectral 
dispersion of 0.75\,\AA{}/pixel. The use of the mosaic detector 
with a  pixel size of 15\,$\mu$m allowed us to cover the
spectral range from 3250 to 6215\,\AA{}. During the run in 2010,  due to its faintness,
the star BAT99\,7 was observed with a 
$0.6\arcsec$ slit and a binning of 2 along the wavelength axis, which results in a spectral resolving 
power of 670. The other stars (WR\,6, WR\,7, WR\,18, and WR\,23) are galactic WR stars and were observed with a 
$0.5\arcsec$ slit ($R\sim1600$). Three of these fours stars are of subtype WN4 and are believed to be single 
stars. WR\,23 is of subtype WC6.

The respective phases for the twelve observations of WR\,6 obtained in service mode in 2011--2012 
are presented in Table~\ref{tab:phase}, including the single observation 
obtained in 2010. For these observations, a slit width of $0.4\arcsec$ was used, resulting in a spectral resolving power 
of $\sim 2000$. Unfortunately, the observation of circular polarization on the night 2011 October 11 
had a very low SNR and could not be used for the measurements of the magnetic field.

The circular polarization observations were carried out using 
the quarter-wave plate at the positions $-45^{\circ}$ and $+45^{\circ}$, whereby the sequence 
$-45^{\circ}$, $+45^{\circ}$, $+45^{\circ}$, $-45^{\circ}$, $-45^{\circ}$, $+45^{\circ}$, \dots{} was adopted
to minimize the cross-talk effect and to cancel errors from 
different transmission properties of the two polarised beams. Moreover, the reversal of the quarter wave 
plate compensates for fixed errors in the relative wavelength calibrations of the two
polarised spectra.

From the raw FORS\,2 data, the parallel and perpendicular beams
are extracted using a pipeline written in the MIDAS environment
by T.~Szeifert, the very first FORS instrument scientist.
This pipeline reduction by default includes background subtraction.
A unique wavelength calibration frame is used for each night.

A first description of the assessment of the longitudinal magnetic field
measurements using FORS\,1/2 spectropolarimetric observations was presented 
in our previous work (e.g., \citealt{Hubrig2004a,Hubrig2004b}, and references therein).
A full description of the currently updated data 
reduction and analysis will be presented in a separate paper (Sch\"oller et 
al., in preparation).
The $V/I$ spectrum is calculated using:

\begin{equation}
\frac{V}{I} = \frac{1}{2} \left\{ 
\left( \frac{f^{\rm o} - f^{\rm e}}{f^{\rm o} + f^{\rm e}} \right)_{-45^{\circ}} -
\left( \frac{f^{\rm o} - f^{\rm e}}{f^{\rm o} + f^{\rm e}} \right)_{+45^{\circ}} \right\}
\end{equation}

\noindent
where $+45^{\circ}$ and $-45^{\circ}$ indicate the position angle of the
retarder waveplate and $f^{\rm o}$ and $f^{\rm e}$ are the ordinary and
extraordinary beams, respectively.  Rectification of the $V/I$ spectra was
performed in the way described by \citet{Hubrig2014}. 
Null profiles $N$ are calculated as pairwise differences from all available 
$V$ profiles.  From these, 3$\sigma$-outliers are identified and used to clip 
the $V$ profiles.  This removes spurious signals, which mostly come from cosmic
rays, and also reduces the noise. 

The strategy for detecting magnetic fields in this work is to look for correlations
between $V/I$ and a profile-dependent diagnostic that would be expected to
yield $V/I$ if magnetic fields are present.
Since the noise in the data precludes testing any specific field model, we note
that for a field with constant $B_{\rm z}$ everywhere, 
we have the relation \citep{angel1970}  

\begin{eqnarray} 
\frac{V}{I} = -\frac{g_{\rm eff}\, e \,\lambda^2}{4\pi\,m_{\rm e}\,c^2}\,
\frac{1}{I}\,\frac{{\rm d}I}{{\rm d}\lambda} \left<B_{\rm z}\right>\, ,
\label{eqn:vi}
\end{eqnarray} 

\noindent 
where $V$ is the Stokes parameter that measures the circular polarization, $I$
is the intensity in the unpolarized spectrum, $g_{\rm eff}$ is the average effective
Land\'e factor, $e$ is the electron charge, $\lambda$ is the wavelength,
$m_{\rm e}$ is the electron mass, $c$ is the speed of light, 
${{\rm d}I/{\rm d}\lambda}$ is the wavelength derivative of Stokes~$I$, and 
$\left<B_{\rm z}\right>$ is the mean longitudinal (line-of-sight) magnetic field.

A field with constant $B_{\rm z}$ is only schematic; it is not intended as a quantification
of the actual average $\left <B_{\rm z} \right >$, merely as a way to characterize it via
an equivalent constant-$B_{\rm z}$ field that would produce the correlation in Eq.~(\ref{eqn:vi}).
For a real situation with spatially varying $B_{\rm z}$, the explicit averaging that applies
to $\left <B_{\rm z} \right >$ implied in Eq.~(\ref{eqn:vi}) is only quantitatively defined
in the case of static atmospheres, but certainly no correlation is expected between
$V$ and ${\rm d} I / {\rm d} \lambda$ in the absence of a magnetic field.
As shown in \citet{gayley2010}, what is different for a field entrained in a hypersonic wind is that
a correlation between the wind flow and the magnetic field direction
induces a corresponding correlation between Doppler and Zeeman shifts, 
altering Eq.~(\ref{eqn:vi}) in ways that depend on the field model. 
Since the SNR in our data is not sufficient for creating detailed magnetic field
models, our goal is only to characterize the overall magnitude of the field strength.
Hence, it is sufficient for our purposes to use any expression that gives
accurate results for a spatially constant $B_{\rm z}$, thereby achieving only schematic 
accuracy at a level of roughly a factor~2, when compared to more detailed or physically plausible magnetic field models 
that the data simply lacks the precision to support.
The systematic uncertainty introduced by this approach will be explored below, but we stress that the
magnetic field magnitudes cited here are all subject to significant systematic uncertainties due to the absence of
a detailed magnetic field model, and the error margins quoted only include the observational uncertainties, not the systematic
ones that relate to a rather vague meaning of the average line-of-sight magnetic field $\left <B_{\rm z} \right >$
in a hypersonic wind with a spatially extended line-forming region and a rapidly falling magnetic field strength.
One may thus regard the mean magnetic field being measured as having the meaning of
a spatially constant longitudinal magnetic field \textit{equivalent}; using that to constrain actual magnetic field models
requires choosing a magnetic field model and is beyond the scope of this investigation.


With this caveat in place, this mean longitudinal magnetic field 
was measured in two ways: using the entire spectrum including  
all lines visible in the spectral region covered by FORS\,2 (see e.g.\ \citealt{hamann2006} for 
line identification), and using all 
lines apart from the line \ion{He}{ii} $\lambda$4686, which is formed farthest in the stellar wind.
Featureless spectral regions were not included in the measurements. 

\begin{figure}
\centering
\includegraphics[width=.23\textwidth]{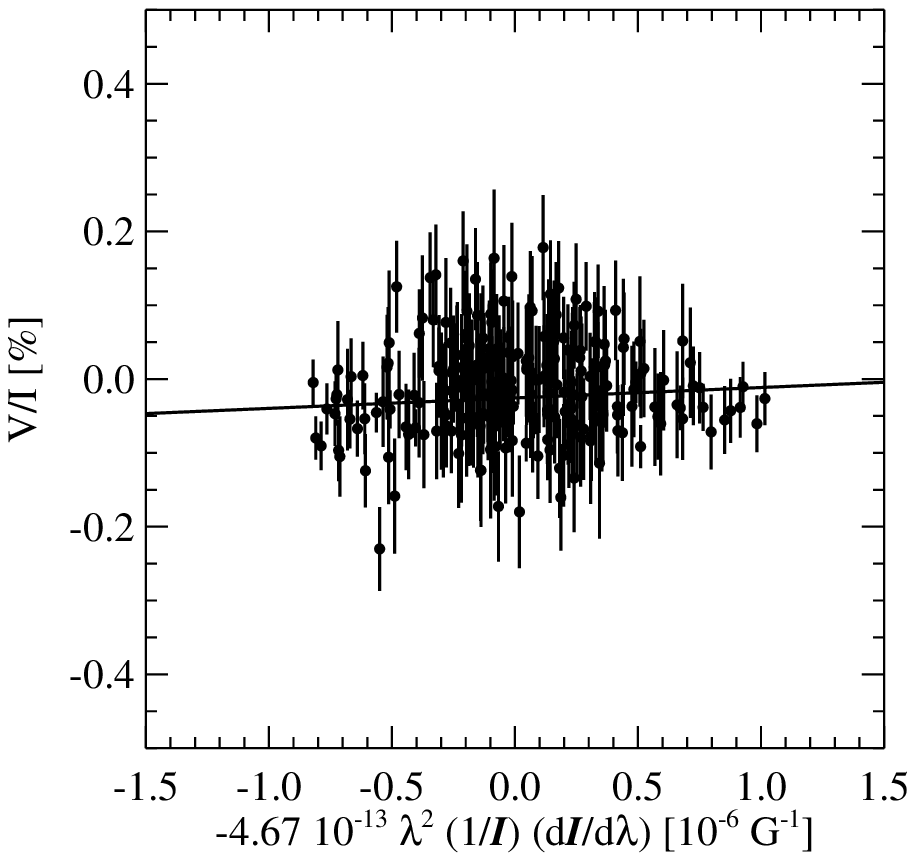}
\includegraphics[width=.23\textwidth]{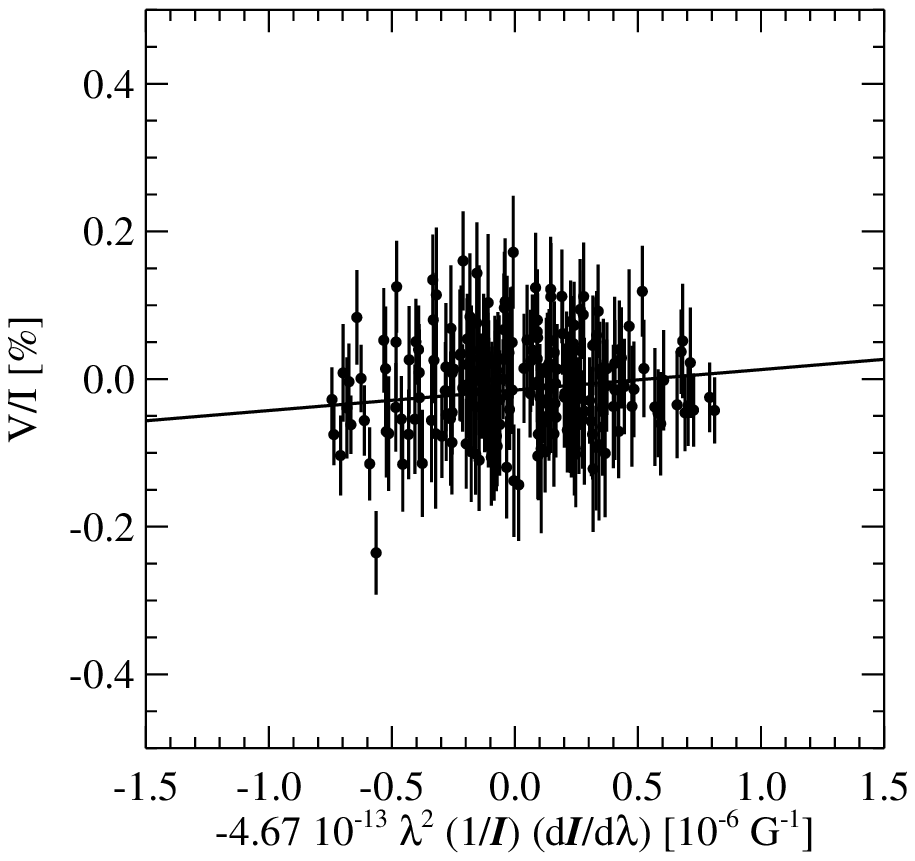}
\caption{
$V/I$ plotted against $-4.67\,10^{-13} \lambda^2 (1/I) ({\rm d}I/{\rm d}\lambda)$.
The solid line shows the linear fit to the data.
The slope of this line translates directly into the $\left<B_{\rm z}\right>$ value.
Due to instrumental polarization, the line might not go through the origin.
Data of WR\,6 from 2010 December 24 for the entire spectrum ({\sl left}) 
and excluding the \ion{He}{ii} $\lambda$4686 line ({\sl right}).
}
\label{fig:phase560}
\end{figure}

\begin{figure}
\centering
\includegraphics[width=.23\textwidth]{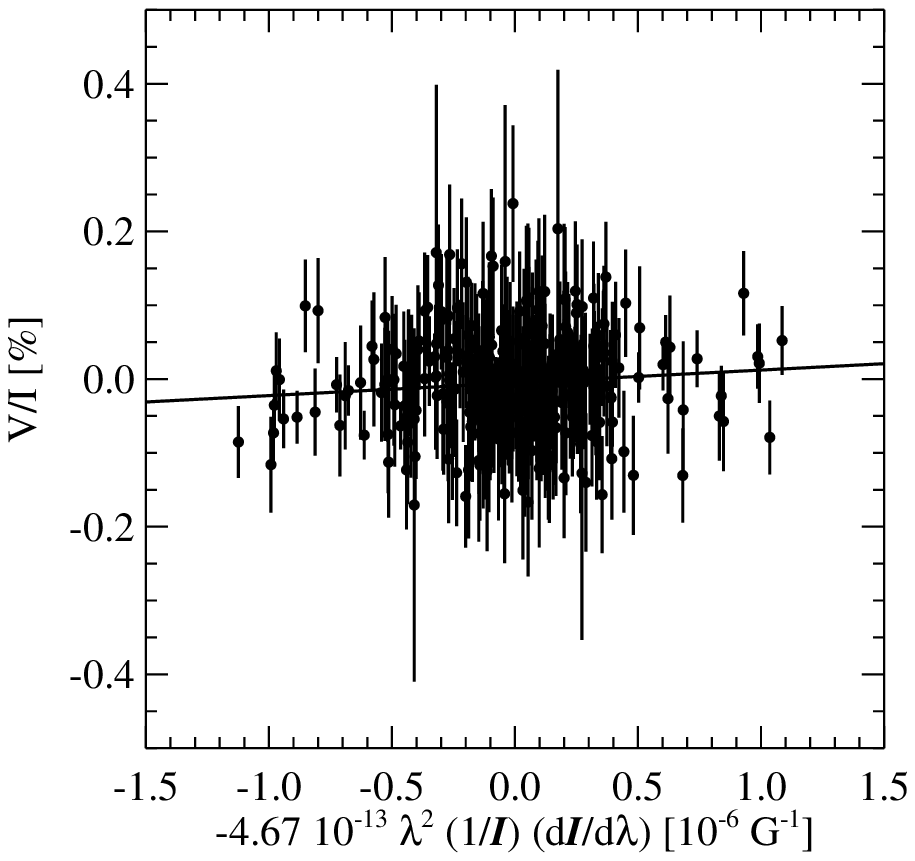}
\includegraphics[width=.23\textwidth]{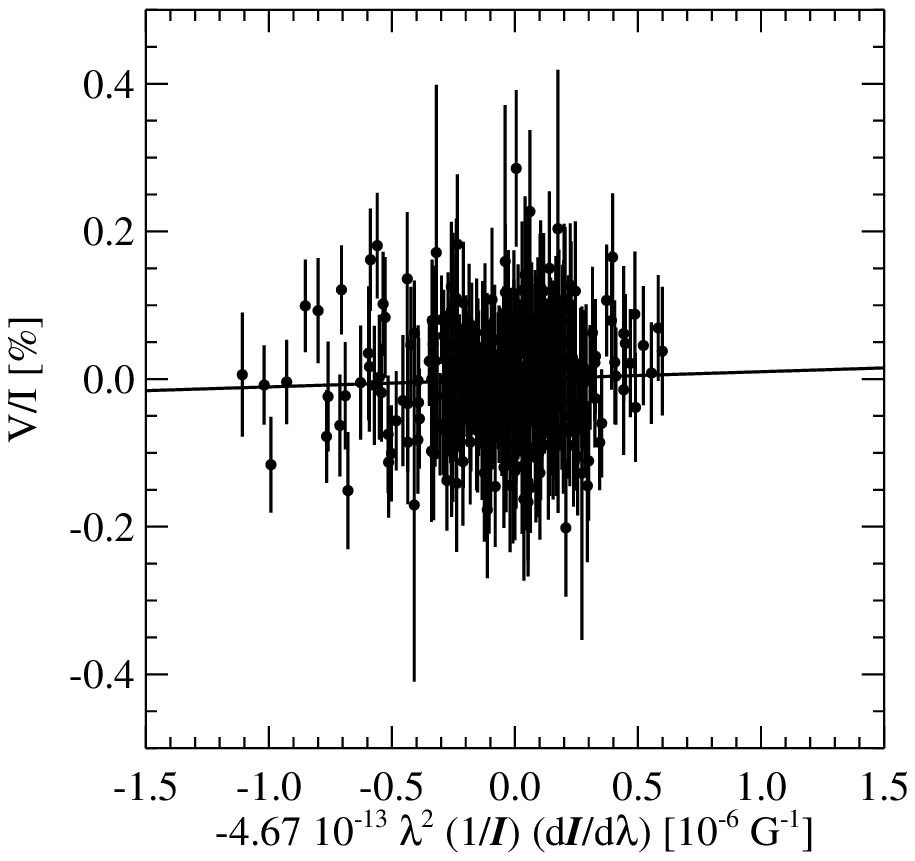}
\caption{
Same as Fig.~\ref{fig:phase560}, but for the night of 2012 January 5.
}
\label{fig:phase643}
\end{figure}

The mean longitudinal magnetic field $\left<B_{\rm z}\right>$ is defined by the slope of the 
weighted linear regression line through the measured data points, where
the weight of each data point is given by the squared SNR
of the Stokes $V$ spectrum. The formal 1$\sigma$ error of 
$\left<B_{\rm z}\right>$ is obtained from the standard relations for weighted 
linear regression. This error is inversely proportional to the rms  
SNR of Stokes $V$. Finally, the factor
$\sqrt{\chi^2_{\rm min}/\nu}$ is applied to the error determined from the 
linear regression, if larger than 1. Furthermore, we have carried out Monte Carlo bootstrapping tests. 
These are most often applied with the purpose of deriving robust estimates of standard errors. 
In these tests, we generate 250\,000 samples from the original data that have the same size as the original data set and analyse 
the distribution of the  $\left<B_{\rm z}\right>$ determined from all these newly generated data sets. 
The measurement uncertainties obtained before and after the Monte Carlo bootstrapping tests were found to be 
in close agreement, indicating the absence of reduction flaws. 
To check the stability 
of the spectral lines along the full sequence of sub-exposures, we have compared 
the profiles of several spectral lines recorded in the parallel beam with the retarder waveplate at $+45^{\circ}$. 
The same was done for spectral lines recorded in the perpendicular beam. 
The line profiles looked identical within the noise.
In Figs.~\ref{fig:phase560} and \ref{fig:phase643}, we show examples for the linear regression of
the plot $V/I$ against $-4.67\,10^{-13} \lambda^2 (1/I) ({\rm d}I/{\rm d}\lambda)$.
The slope of the line fitted to the data directly translates into the value for $\left<B_{\rm z}\right>$.
Fig.~\ref{fig:phase560} shows the results for the night of 2010 December 24, while Fig.~\ref{fig:phase643}
shows the results for the night of 2012 January 5, both for the entire spectrum and excluding the
\ion{He}{ii} $\lambda$4686 line.
The values determined for the magnetic field can be found in Table~\ref{tab:wr6}.

As a circularly polarized standard, we observed on two occasions, on 2010 December~24 and 25,
the strongly magnetic A0p star HD\,94660. HD\,94660 has a longitudinal 
magnetic field that varies about a mean value of $\sim -2$\,kG with an amplitude of a few hundred Gauss 
over a period of 2800\,d \citep{bailey2015}. Both measurements show a $-$2.3\,kG field,
which is fully consistent with the values for the longitudinal magnetic field at the considered rotation phases 
expected from the variation curve defined by \citet{bailey2015}.

Noteworthy, we recently presented a detailed comparison \citep{fossati2015} between our measurements and the independent measurements 
of another group that uses reduction and analysis techniques developed strictly following \citet{bag2012}.
The comparison showed results that agree well within a Gaussian distribution.
While not all 3$\sigma$ detections can be considered as genuine for single observations with FORS\,2,
they appear to be genuine in a number of studies where the measurements show
smooth variations over a rotation period, similar to those found for the magnetic Of?p
stars HD\,148937 and CPD\,$-$28$^{\circ}$\,2561 \citep{Hubrig2008,Hubrig2011,Hubrig2013,Hubrig2015}.
In these studies, not a single reported detection reached a 4$\sigma$ significance level.

\begin{figure}
\centering
\includegraphics[width=.45\textwidth]{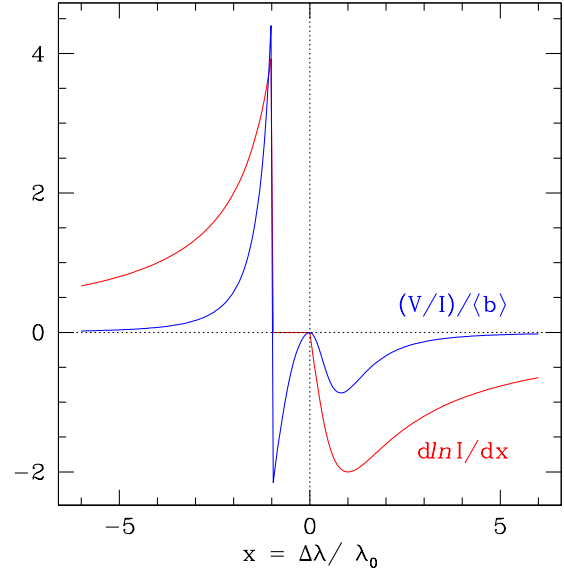}
\caption{Using the expression from  \citet{Gayley2012} for a recombination wind emission line threaded by a split monopole 
magnetic field, $d\ln I/dx$ (red line) and $(V/I)/\left<b\right>$ (blue line) are plotted against the normalized wavelength $x$.  
The case shown is for an optically thick line with $p=6$.
}
\label{fig:rico_1}
\end{figure}

\begin{figure}
\centering
\includegraphics[width=.45\textwidth]{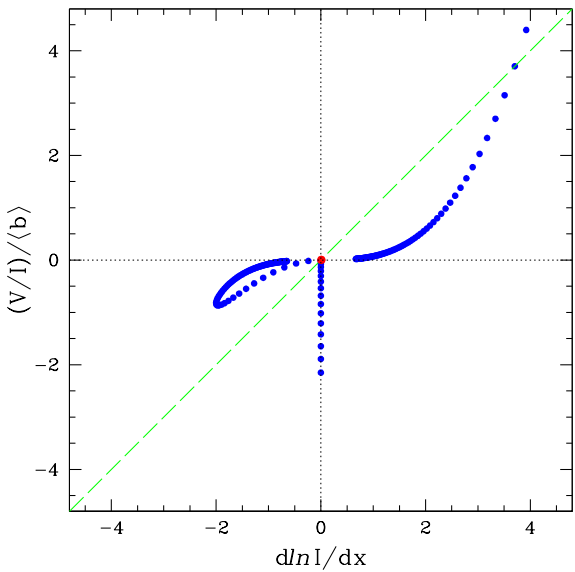}
\caption{Following Fig.~\ref{fig:rico_1}, we present a plot of $(V/I)/\left<b\right>$ against $d\ln I/dx$ (blue points).  The upper 
right quadrant is for points in the line profile with $x<-1$.  The vertical distribution of points located at 
$d\ln I/dx$ corresponds to the flat-top portion of the line shape, for $-1<x<0$.  The points in the lower left quadrant 
are for redshifted wavelengths in the line, with $x>0$.  The dashed green line is the expected relation for a static atmosphere.
}
\label{fig:rico_2}
\end{figure}

Equation~(\ref{eqn:vi}) is usually applied to stars with weak winds and no rapid rotation.
To qualitatively understand the application
of this equation to wind lines in the atmospheres of WR stars, we present in Figs.~\ref{fig:rico_1} 
and \ref{fig:rico_2} the schematic quality of equation~(\ref{eqn:vi}) in winds,
following the model presented in \citet{Gayley2012}.  
In these figures, $x$ is the line width in units of the radial Doppler shift at the wind photosphere.
$p$ sets the power law for the radial falloff of the volume emissivity.
We take $p=6$ to simulate a majority-ion density-squared emissivity in  the 
acceleration zone of a wind with $v \varpropto r$.
The $\left<b\right>$ parameter gives the line-of-sight field averaged over the region in the wind
that contributes to the profile at $x=1$ in the uniformly expanding wind model with a
split monopole field. 
As this is the region where much of the $V$ signal originates,
it provides a reasonable meaning for the average detected field. 
The unit of $\left<b\right>$ is the field necessary to produce a Zeeman shift
equal to the Doppler shift of the radial velocity at the wind photosphere (i.e.,
the Doppler shift corresponding to $x=1$).

From Fig.~\ref{fig:rico_1}, we can see that the detailed shape of $V/I$ is not closely proportional to
$dlnI/d\lambda$, owing to spatial variations in $\left<B_{\rm z}\right>$ in the detailed model used for 
those plots, but the overall magnitude of the nonzero $V/I$ is similar, especially for larger $V/I$ that
would rise more significantly above the noise. 
Moreover, there is no reason to expect any correlation whenever the $V/I$ is purely noise.
Still, one must accept significant systematic uncertainties in the quantitative meaning of an average
longitudial magnetic field in the absence of a specific magnetic field model, so the results quoted are not intended
to be directly compared to mean-field upper limits 
acquired via other means, at a level better than roughly a factor of 2.
For this reason, the conclusion by \citet{chevro2013}
that the mean longitudinal magnetic field is below about 100\,G is not
necessarily in conflict with our results, as we see evidence of magnetic fields
at the same level as their upper bound (see discussion in Sect.~\ref{sect:meas}).

Linear polarization observations in service mode were obtained for eight angles of the 
retarder waveplate: $0.0^{\circ}$, $22.5^{\circ}$, $45.0^{\circ}$, $67.5^{\circ}$, 
$90.0^{\circ}$, $112.5^{\circ}$, $135.0^{\circ}$, $157.5^{\circ}$. During the visitor run in 2010,
only the first four positions of the half-wave plate were used.  

A net linear polarization from Thomson scattering and line scattering is produced when spherical symmetry is broken. 
Existing observations revealed that the linear polarization is quite moderate, 
but, of course, nothing is known about linear polarization for most of the targets in our sample.
We note that the linear polarization is of immediate scientific interest itself, as the data can be interpreted not only with 
respect to deviations from spherical symmetry,
but -- if variable -- also with respect to the temporal evolution of wind 
structures.

The data reduction for observations of linear polarization was done with the ESO pipeline (v4.9.18) 
recipes (fors\_pmos\_calib, fors\_pmos\_science). 
For all targets, we calculated the Stokes parameters $P_Q = Q/I$ and $P_U = U/I$ 
as defined in \citet{shurcliff62}.

To calculate the $P_Q$ and $P_U$ parameters, we used the
equations described in the FORS\,1/2 User Manual and on the ESO FORS\,2 web page 
(http://www.eso.org/\linebreak[1]sci/\linebreak[1]facilities/\linebreak[1]paranal/\linebreak[1]instruments/\linebreak[1]fors/\linebreak[1]inst/\linebreak[1]pola.html):

\begin{eqnarray*}
\textstyle
P_Q &
\textstyle
= &
\textstyle
\frac{1}{2} \left\{ \left( \frac{f^o-f^e}{f^o+f^e} \right)_{\alpha=0^\circ} - \left( \frac{f^o-f^e}{f^o+f^e} \right)_{\alpha=45^\circ} \right\} \\
 &
\textstyle
+ &
\textstyle
\frac{1}{2} \left\{ \left( \frac{f^o-f^e}{f^o+f^e} \right)_{\alpha=90^\circ} - \left( \frac{f^o-f^e}{f^o+f^e} \right)_{\alpha=135^\circ} \right\}
\end{eqnarray*}
\begin{eqnarray*}
\textstyle
P_U &
\textstyle
= &
\textstyle
\frac{1}{2} \left\{ \left( \frac{f^o-f^e}{f^o+f^e} \right)_{\alpha=22.5^\circ} - \left( \frac{f^o-f^e}{f^o+f^e} \right)_{\alpha=67.5^\circ} 
\right\} \\
 &
\textstyle
+ &
\textstyle
\frac{1}{2} \left\{ \left( \frac{f^o-f^e}{f^o+f^e} \right)_{\alpha=112.5^\circ} - \left( \frac{f^o-f^e}{f^o+f^e} \right)_{\alpha=157.5^\circ} \right\}
\end{eqnarray*}

The reduction procedure consists of bias
subtraction, flat fielding, extraction, and wavelength calibration of
the spectra. 
The total linear polarization is determined as $P_L =\sqrt{P_Q^2+P_U^2}$ 
and the polarization position angle as $\theta_P= \frac{1}{2}$ arctan$\left(P_U/P_Q\right)$.
The associated errors for the determination of $P_Q$, $P_U$,  $P_L$ and $\theta_P$ 
are estimated from error propagation, based on pure photon noise in the raw data.
They are usually of the order of 0.02--0.09\% for $P_Q$, $P_U$ and $P_L$, 
and of about 0.4--1.3$^\circ$ for the determination of $\theta_P$. 

Since FORS\,2 is mounted in the UT\,1 Cassegrain focus, its spectropolarimetric calibration 
is relatively stable. Instrumental polarization measured using unpolarized standard stars is rather low,
less than 0.2\%. Our observations of the 
spectropolarimetric standard NGC\,2024-1 revealed the values $P_L = 9.58\pm0.20$\%
and $\theta_P = 136.61\pm0.42^\circ$, which are in 
very good agreement with the previous measurements $P_L = 9.65\pm0.12$\% and 
$\theta_P = 136.82\pm0.34^\circ$ by \citet{ignace09}.

\section{Results of the magnetic field measurements}
\label{sect:meas}

\begin{table*}
\caption[]{
Longitudinal magnetic field measurements for the entire spectra  ($\left<B_{\rm z} \right>_{\rm all}$) 
and excluding the \ion {He}{ii} $\lambda$4686 line ($\left<B_{\rm z} \right>_{-4686}$)
for the WR stars observed in 2010. 
$\left<N_{\rm z} \right>$ values  are calculated using the null spectra.
All quoted errors are 1$\sigma$ uncertainties.
}
\centering
\begin{tabular}{llr r@{$\pm$}l r@{$\pm$}l r@{$\pm$}l}
\hline
\hline 
\noalign{\vspace{1mm}}
Object & 
\multicolumn{1}{c}{MJD} &
\multicolumn{1}{c}{SNR$_{\rm cont}$} &
\multicolumn{2}{c}{$\left<B_{\rm z} \right>_{\textrm{all}}$} & 
\multicolumn{2}{c}{$\left<B_{\rm z} \right>_{-4686}$} &
\multicolumn{2}{c}{$\left<N_{\rm z} \right>$} \\
 &
 &
 &
\multicolumn{2}{c}{[G]} &
\multicolumn{2}{c}{[G]} &
\multicolumn{2}{c}{[G]} \\
\noalign{\vspace{1mm}}
\hline
\noalign{\vspace{1mm}}
BAT99\,7 & 55554.1024 & 1318 &     37 & 92  &    327 & 141  &     66 & 78  \\  
BAT99\,7 & 55555.0710 & 1131 &  $-$64 & 79  &     24 & 124  &    142 & 65  \\
WR\,6    & 55555.1577 & 1353 &    131 &  68 &    258 &  78  &    11 &  50 \\
WR\,7    & 55554.2477 & 1028 &  $-$65 & 46  &  $-$60 & 49   &   $-$1 & 46  \\
WR\,7    & 55555.2047 & 1302 &  $-$49 & 39  &     15 & 42   &     10 & 37  \\
WR\,18   & 55555.2900 & 1385 &  $-$86 & 42  &  $-$76 & 43   &  $-$19 & 47  \\
WR\,23   & 55554.3245 & 1600 &  $-$35 & 34  & $-$120 & 48   &      2 & 28  \\
WR\,23   & 55555.3440 & 994  &  $-$87 & 40  &  $-$80 & 81   &     65 & 44  \\
\hline         
\label{tab:bfield_2010_1}
\end{tabular}  
\end{table*} 

The results of our magnetic field measurements using FORS\,2 polarimetric spectra obtained
in 2010, both for the entire spectrum and for the 
measurements carried out excluding the strongest emission line belonging to \ion{He}{ii} $\lambda$4686
are presented in Table~\ref{tab:bfield_2010_1},
where we also collect information on the modified Julian date, the SNR,
and the measurements obtained using the null spectra.

Trying to explain 
the huge rotational velocity of about 160\,km\,s$^{-1}$ at the stellar surface of BAT99\,7, 
\citet{shenar2014} suggested the existence of a very strong photospheric magnetic field of up to 32\,kG.
The magnetic field of this star was measured on two consecutive nights in 2010, but no detection at a significance
of 3$\sigma$ was achieved in either measurement.
The magnetic field, if present, would likely be variable. The highest value for the longitudinal magnetic field, 
$\left<B_{\rm z} \right>_{-4686} = 327\pm141$\,G at a significance level of 2.3$\sigma$, was measured on the first night.
On the second night, our measurements result in the non-detection $\left<B_{\rm z} \right>_{-4686} = 24\pm124$\,G.
Certainly, future monitoring of the magnetic field in this star will be worthwhile,
to see if there is a field present that is undergoing modulation from rapid rotation, 
or if the observation is simply random noise.

For WR\,23, we measure on the first
night $\left<B_{\rm z} \right>_{-4686} = -120\pm48$\,G at a significance level of 2.5$\sigma$ and  on the second night
$\left<B_{\rm z} \right>_{-4686} = -80\pm81$\,G. For the  
stars WR\,18 and WR\,7, magnetic fields are measured at a significance level of about 2$\sigma$ and below.

Among the stars in our sample, only the star WR\,6 was reported to show the existence of 
structures called CIRs, which are believed to be created by density contrasts owing to velocity
shear between fast and slow wind streaklines bent by rotation (e.g.\ \citealt{cranmer1996}).
In some studies also magnetic spots have been invoked as the likely cause of CIRs in hot-star winds.
However, previous attempts detected no such fields above
several 100\,G (e.g., \citealt{kholtygin2011}; \citealt{chevro2013}). 

The magnetic field with the highest significance level in our study, with $\left<B_{\rm z} \right>_{-4686} = 258\pm78$\,G,
was detected in WR\,6 during our visitor run in 2010.
A first direct search for a magnetic field via the circular polarization of Zeeman 
splitting in this star was recently carried out by \citet{chevro2013} using 
the ESPaDOnS spectropolarimeter at the Canada-France-Hawaii Telescope. No magnetic field was unambiguously detected 
by this team. Assuming that the star is intrinsically magnetic and can be described by a split monopole configuration, the 
authors found an upper limit
on the order of 100\,G for the strength of its magnetic field in the line-forming regions of the stellar wind.

\begin{table}
\caption[]{
Longitudinal magnetic field measurements for the entire spectra ($\left<B_{\rm z} \right>_{\textrm{all}}$) 
and excluding the \ion {He}{ii} $\lambda$4686 line ($\left<B_{\rm z} \right>_{-4686}$)
for the star WR\,6 observed in 2011--2012. 
$\left<N_{\rm z} \right>$ values  are calculated using the null spectra.
All quoted errors are 1$\sigma$ uncertainties.
The observation on the night of 2011 October 11 had
very low SNR and could not be used for the measurement of the magnetic field.  
}
\centering
\begin{tabular}{llr r@{$\pm$}l r@{$\pm$}l r@{$\pm$}l}
\hline
\hline 
\noalign{\vspace{1mm}}
\multicolumn{1}{c}{\textrm{MJD}} &
\multicolumn{1}{c}{rotation} &
\multicolumn{1}{c}{rotation} &
\multicolumn{2}{c}{$\left<B_{\rm z} \right>_{\textrm{all}}$} &
\multicolumn{2}{c}{$\left<B_{\rm z} \right>_{-4686}$} &
\multicolumn{2}{c}{$\left<N_{\rm z} \right>$} \\
 &
\multicolumn{1}{c}{phase} &
\multicolumn{1}{c}{cycle} &
\multicolumn{2}{c}{[G]} &
\multicolumn{2}{c}{[G]} &
\multicolumn{2}{c}{[G]} \\
\noalign{\vspace{1mm}}
\hline
\noalign{\vspace{1mm}}
55555.1577 & 0.560 & $-$77& 131 &  68   &    258 &  78   &    11 &  50 \\       
55845.2716 & 0.596 &    0 &    70 &  56   &     86 &  62   & $-$20 &  64 \\       
55880.1638 & 0.861 &    9 & $-$12 &  54   &  $-$52 &  62   & $-$13 &  51 \\       
55903.3172 & 0.009 &   15 &$-$100 &  52   & $-$126 &  65   &  $-$2 &  83 \\       
55906.1257 & 0.754 &   16 &    30 &  52   &     19 &  61   &    50 &  58 \\       
55909.2102 & 0.573 &   16 &   35 &  59   &     73 &  70   & $-$20 &  56 \\       
55928.0760 & 0.583 &   21 &   64 &  76   &     62 &  77   &    38 &  75 \\       
55929.0803 & 0.849 &   22&    37 &  68   &  $-$16 &  73   &    28 &  61 \\       
55932.0686 & 0.643 &   23&   171 &  56   &    111 &  61   &  $-$9 &  63 \\       
55933.0511 & 0.904 &   23&     4 &  59   &      9 &  68   &    17 &  66 \\       
55934.0611 & 0.172 &   23&    62 &  43   &     61 &  44   &    41 &  63 \\       
55935.2491 & 0.487 &   23&   128 & 106   &    144 &  90   &    13 &  44 \\    
\hline         
\label{tab:wr6}
\end{tabular}  
\end{table} 

The results of all magnetic field measurements of WR\,6 using FORS\,2 polarimetric spectra obtained
in visitor and service mode, both for the entire spectrum and for the 
measurements carried out excluding the very strong emission line belonging to \ion{He}{ii} $\lambda$4686
are presented in Table~\ref{tab:wr6},
where we also give information on the modified Julian date,
the rotation phase, and the measurements obtained using null spectra. 
The phase of each
observation was obtained from the mid-times of exposure and the
ephemeris given by \citet{lamo1986}, based on long-term photometric monitoring
($HJD=2446153.61 + 3.766E$).
We note  that the zero point of the phase in our measurements is arbitrary since about 25 years 
have passed between \citeauthor{lamo1986}'s latest data and those presented here for WR\,6.

\begin{figure}
\centering
\includegraphics[width=.45\textwidth]{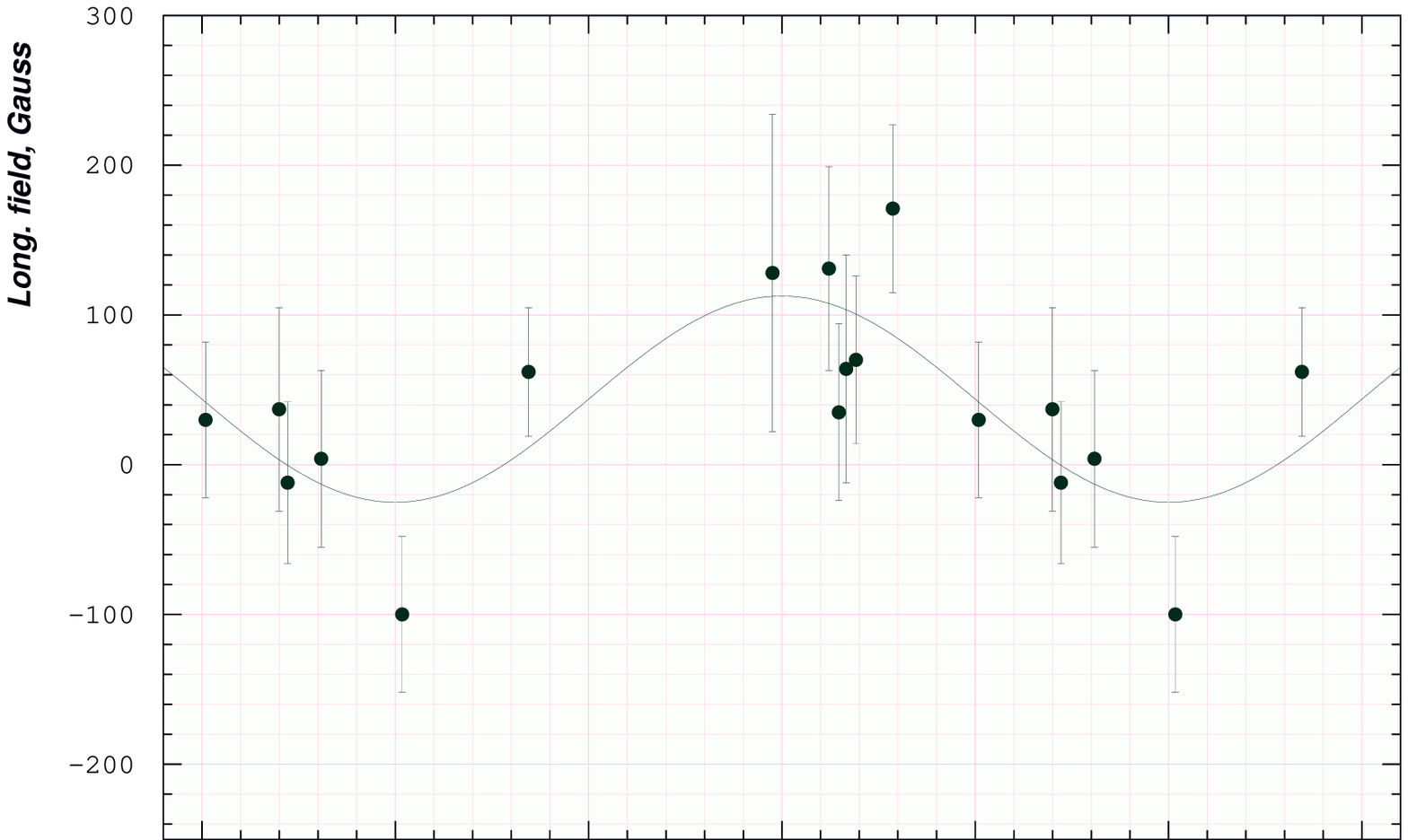}
\includegraphics[width=.45\textwidth]{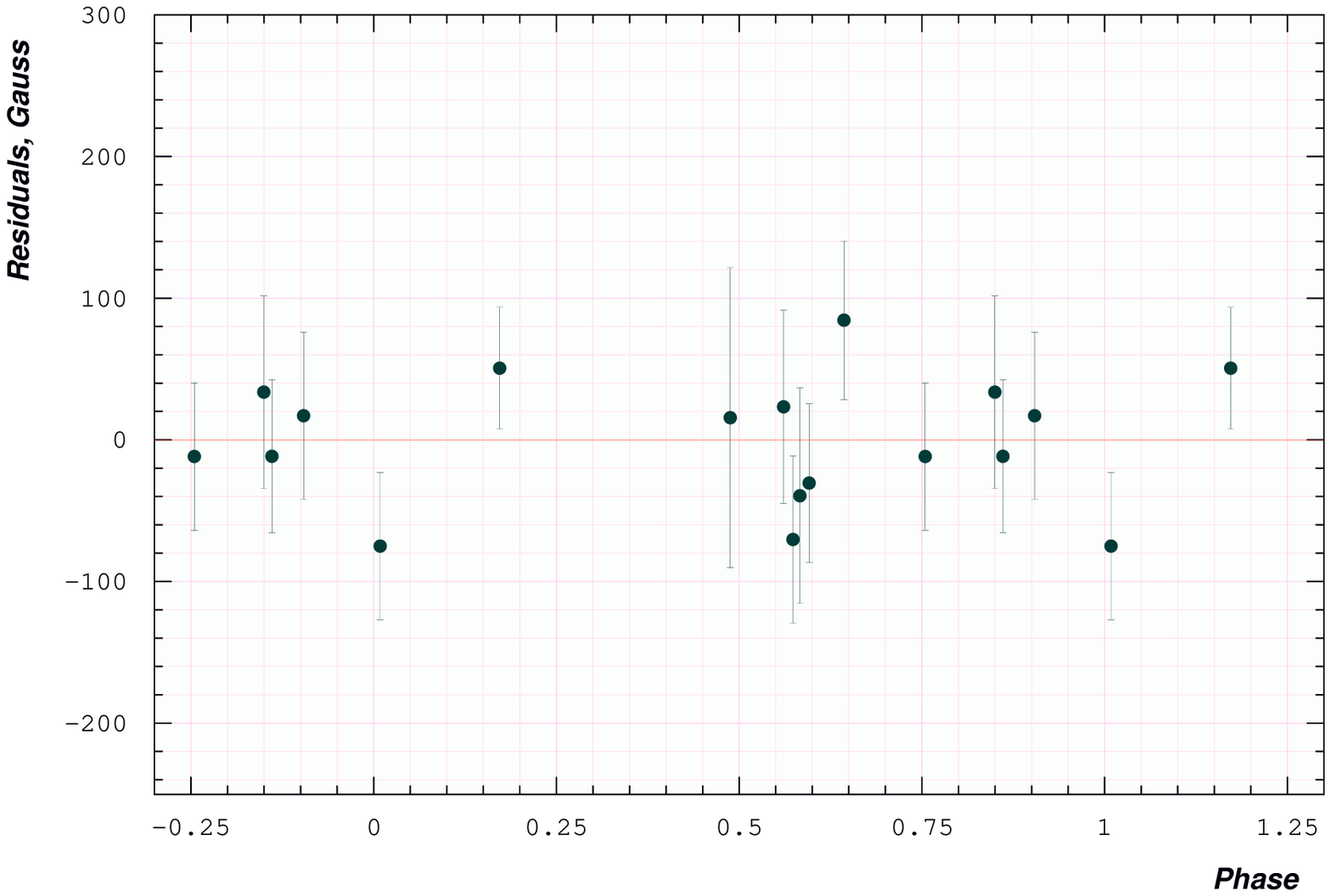}
\caption{Phase diagram and its best sinusoidal fit for the mean longitudinal magnetic field 
measurements of WR\,6 using the entire spectrum. The residuals (observed -- calculated) are shown in the lower 
panel. The deviations are mostly of the same order as the error bars, and no systematic trends are obvious, 
which justifies a single sinusoid as a fit function.
The sinusoid can be described by an offset of $\overline{\left< B_{\rm z}\right>}= 44\pm15$\,G
and an amplitude of $A_{\left< B_{\rm az}\right>}=69\pm21$\,G.
}
\label{fig:spectrum}
\end{figure}

\begin{figure}
\centering
\includegraphics[width=.45\textwidth]{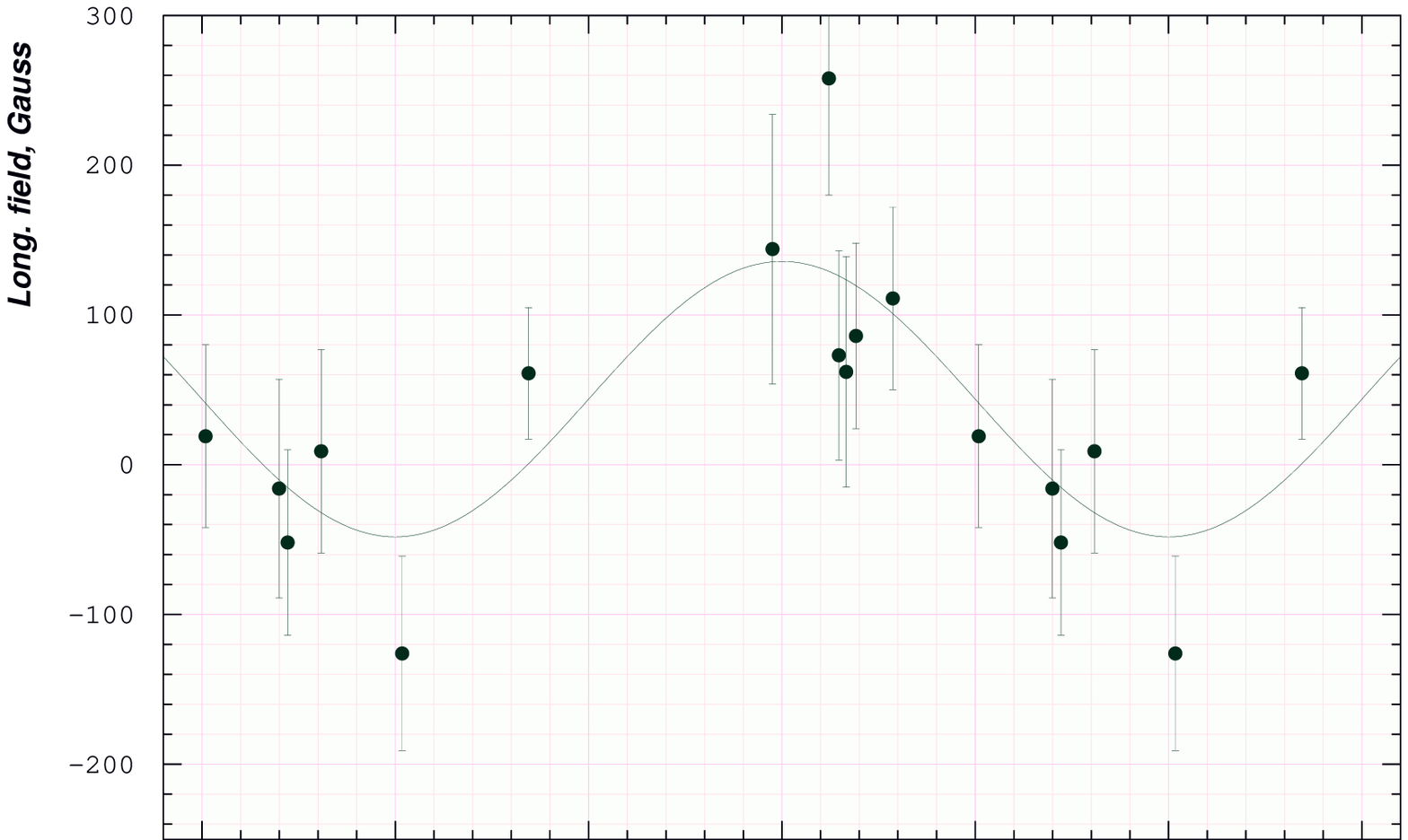}
\includegraphics[width=.45\textwidth]{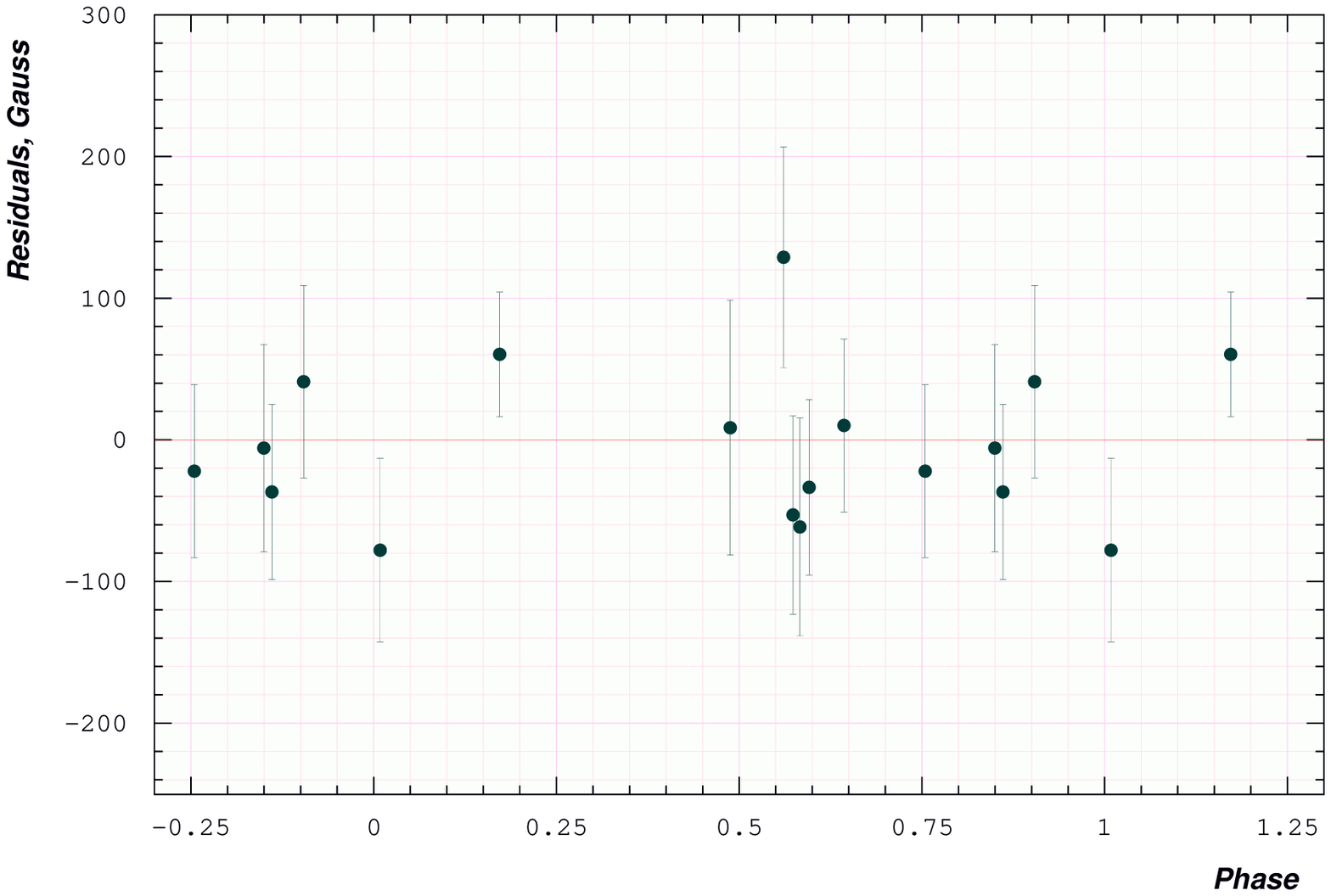}
\caption{Phase diagram and its best sinusoidal fit for the mean longitudinal magnetic field 
measurements of WR\,6 excluding the \ion{He}{ii} $\lambda$4686 line. The residuals (observed -- calculated) 
are shown in the lower panel. The deviations are mostly of the same order as the error bars, and 
no systematic trends are obvious, which justifies a single sinusoid as a fit function.
The sinusoid can be described by an offset of $\overline{\left< B_{\rm z}\right>}= 44\pm17$\,G
and an amplitude of $A_{\left< B_{\rm az}\right>}=92\pm24$\,G.
}
\label{fig:spectrum_4686}
\end{figure}

\begin{figure}
\centering
\includegraphics[angle=0,width=.45\textwidth]{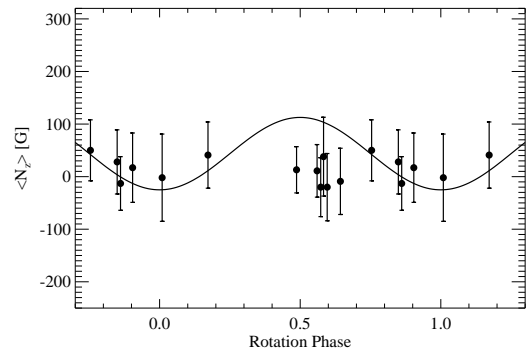}
\caption{Mean longitudinal magnetic field measurements using the null spectra. The overplotted sine curve corresponds
to the sinusoidal fit presented in Fig.~\ref{fig:spectrum}.}
\label{fig:spectrum_null}
\end{figure}

The measurements phased with this rotation period  using the entire spectrum including all available
lines and excluding the \ion{He}{ii} $\lambda$4686 line, and the best sinusoidal fits calculated for these measurements,
taking into account the already known rotation period $P_{\rm rot} = 3.766$\,d, are
presented in Figs.~\ref{fig:spectrum} and \ref{fig:spectrum_4686}, respectively. 
In Fig.~\ref{fig:spectrum_null},
we present the mean longitudinal magnetic field measurements using the null spectra. 
As is shown in this figure, no similar variability of the measured field values using the null spectra is 
detected over the rotation period. The distribution of the measured values
appears random and does not show any trend with the rotation phase.

The field is reversing with the negative and positive 
field extrema appearing at the rotation phases around 0 and 0.5, respectively.
Only two measurements, namely those obtained at the rotation phases of 0.560 and 0.643, 
reveal statistically significant values of the longitudinal magnetic field at a significance level of 
3.3$\sigma$ ($\left<B_{\rm z} \right>_{-4686}=258\pm78$\,G) and 3.1$\sigma$ ($\left<B_{\rm z} \right>_{all}=171\pm56$\,G).
A few more measurements show a significance level of about 2$\sigma$. The presence
of measurements at lower significance levels in some rotation phases is expected 
due to the sinusoidal character of the field behaviour with reversing polarities. 

From the sinusoidal fits to our data, we obtain 
for the measurements using the entire spectrum a mean value for the variable longitudinal magnetic field
$\overline{\left< B_{\rm z}\right>}= 44\pm15$\,G and an amplitude of the field variation 
$A_{\left< B_{\rm az}\right>}=69\pm21$\,G. 
Using the entire spectrum without the \ion{He}{ii} $\lambda$4686 line, we obtain 
$\overline{\left< B_{\rm z}\right>}= 44\pm17$\,G and $A_{\left< B_{\rm az}\right>}=92\pm24$\,G.
For the best fitting models superimposed on the data presented in Figs.~\ref{fig:spectrum} and 
\ref{fig:spectrum_4686}, we calculate a reduced $\chi^2$ of 0.83 and 0.84, respectively. 
In  contrast, the reduced $\chi^2$ values calculated assuming a model in which the longitudinal magnetic
field is constant and equal to 0, is 1.56 for the measurements including the \ion{He}{ii}~4686 line and 1.92
for the measurements excluding the \ion{He}{ii}~4686 line. Noteworthy, we obtain $\chi^2=0.44$
for the zero-field model, if we use 
the $\left<N_{\rm z} \right>$ values obtained using the null spectra.

From the observed sinusoidal modulation, we learn that the magnetic field structure
shows two poles and a symmetry axis, tilted with respect to the rotation axis.
The simplest model for this magnetic field geometry is based on the assumption that the studied stars 
are oblique dipole rotators,
i.e., their magnetic field can be approximated by a dipole with the magnetic axis 
inclined to the rotation axis.
Of course the strong wind would be expected to extensively modify any simple field
geometry, but the precision of the data is insufficient to explore the field in detail.

The angle of obliquity of the magnetic axis $\beta$ is constrained by

\begin{equation} 
r = \frac{\left< B_{\rm z}\right>^{\rm min}}{\left< B_{\rm z}\right>^{\rm max}} 
  = \frac{\cos \beta \cos i - \sin \beta \sin i}{\cos \beta \cos i + \sin \beta \sin i}, 
\end{equation} 
 
\noindent 
so that the obliquity angle $\beta$ is given by
 
\begin{equation} 
\beta =  \arctan \left[ \left( \frac{1-r}{1+r} \right) \cot i \right]. 
\label{eqn:4} 
\end{equation}

However, because of the dense winds in WR stars, for most of them measuring the projected rotation velocity
$v\,\sin\,i$ is not possible, i.e.\ the rotation axis inclination $i$ remains undefined.
Using an estimate of the stellar radius $R= 2.65\,R_\odot$ for WR\,6 \citep{hamann2006} and
the rotation period $P_{\rm rot} = 3.766$\,d,  
we obtain $v_{\rm eq}=35.6$\,km\,s$^{-1}$.
Since the limb-darkening is also unknown for WR stars,
we can only assume that $B_{\rm d} \ge 3\,\left<B_{\rm z}\right>_{\rm max}$
resulting in a minimum dipole strength of $\sim$300--400\,G.
We note that earlier marginal detections in a few WR stars 
were determined in the framework of a split monopole magnetic field chosen  because  of its  generic
character  in  a  dense,  radial  stellar  wind  \citep{gayley2010}. On the other hand, 	
\citet{chevro2014} suggest in their work that the comparison  
between theoretical predictions and observations should  be considered with care.  

Based on the presented measurements, we suggest a lower limit for the magnetic field in the line-forming regions of the 
stellar wind $B_{\rm wind}\sim300$\,G. Since different emission lines form at different distances from the
stellar surface and cover different zones of the wind, they sample different field strengths.
Knowing roughly the width of the lines, combined with the wind velocity-law index $\beta$, it is possible to 
approximate the surface field strength, assuming $1/r^2$ scaling, where $r = r(v)$. Following the considerations
on magnetic and line-emission zone parameters in WR\,6 by \citet{chevro2013}, we estimate 
that the surface value of the magnetic field is on the order of a few kiloGauss.

\begin{figure}
\centering
\includegraphics[angle=0,width=.45\textwidth]{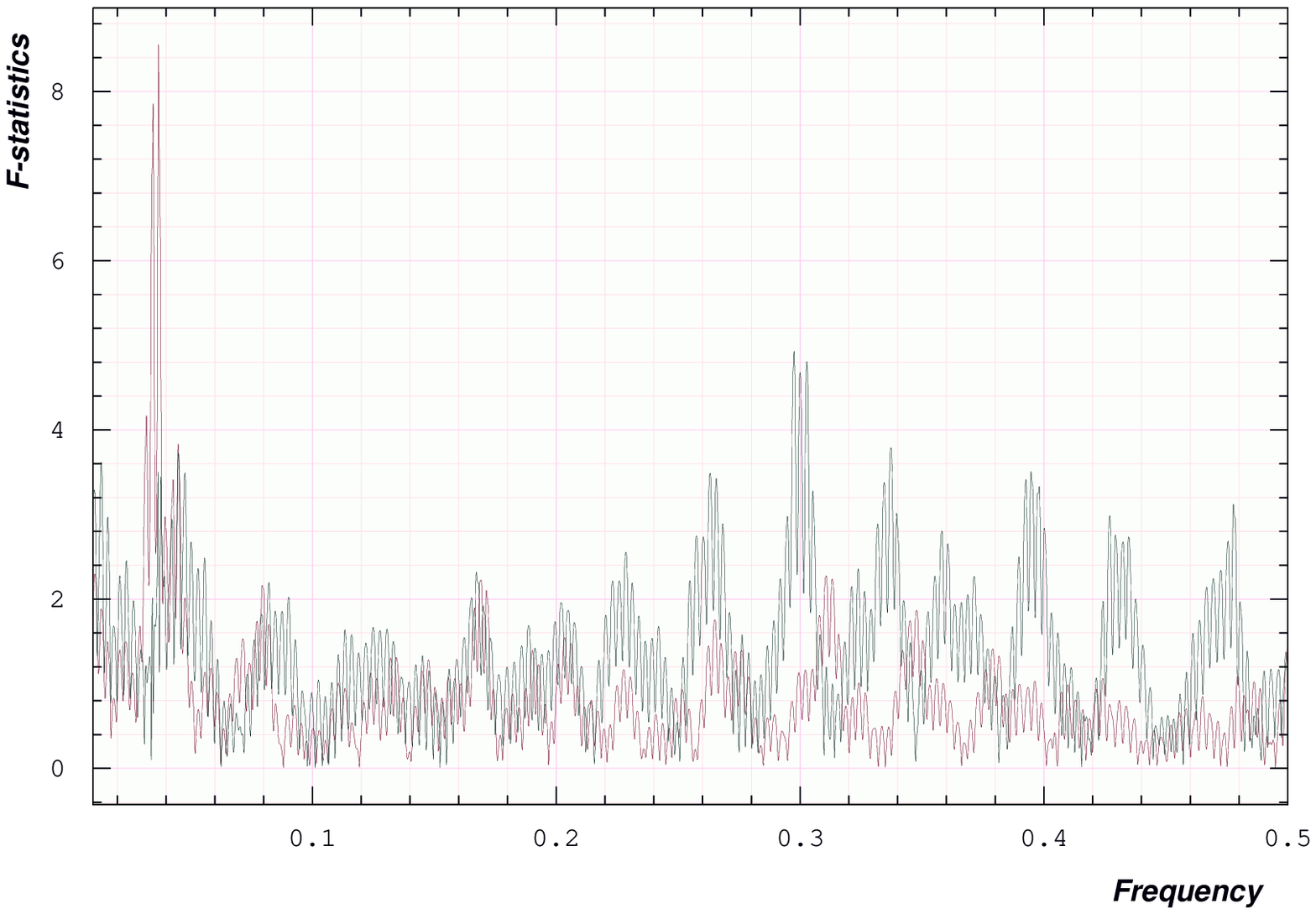}
\includegraphics[angle=0,width=.45\textwidth]{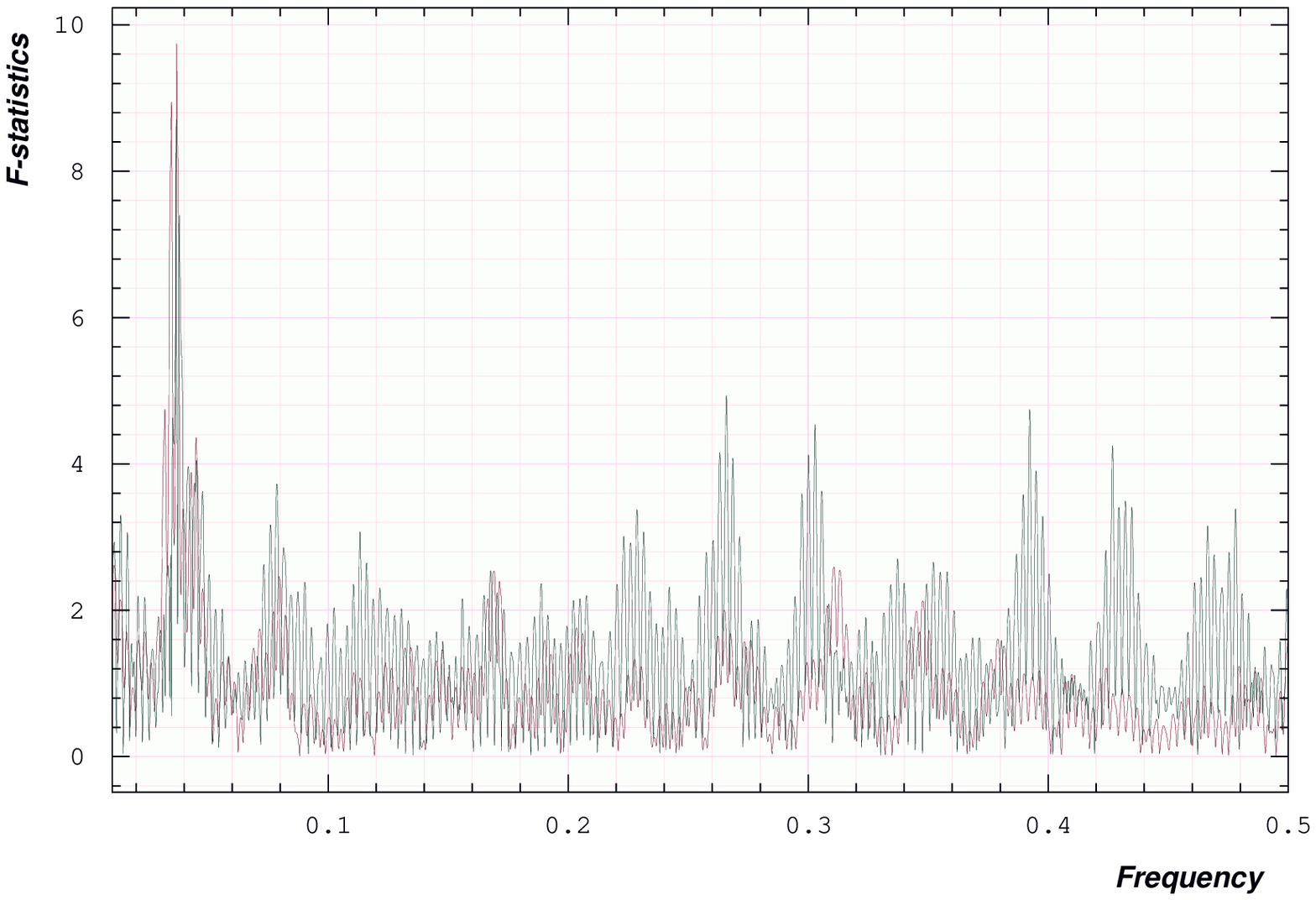}
\caption{Frequency  periodograms (in d$^{-1}$) for longitudinal magnetic field measurements using the 
entire spectrum (upper panel) and 
for the measurements carried out excluding the \ion{He}{ii} $\lambda$4486 line (lower panel).
The window function is indicated by the red color. 
}
\label{fig:periodo}
\end{figure}

Interestingly, in spite of the rather low number of magnetic field measurements, the presence of rotation-modulated 
magnetic variability is also indicated in our frequency periodograms presented in Fig.~\ref{fig:periodo}.
The periodogram shown in the upper panel was obtained for the measurements using the entire 
spectrum, while that in the lower panel was obtained for the measurements carried out excluding 
the \ion{He}{ii} $\lambda$4486 line. 
A frequency analysis was performed using a non-linear least  squares
fit to the multiple harmonics utilizing the Levenberg-Marquardt
method \citep{press1992} with the optional possibility of prewhitening the trial harmonics. To detect 
the most probable period, we  calculated  the  frequency spectrum and for each trial
frequency we performed a statistical F-test of the null hypothesis for the absence of periodicity 
\citep{seber1977}.
The resulting F-statistics can be thought of as the total sum including covariances of the ratio 
of harmonic amplitudes to their standard deviations, i.e.\ an SNR.
The highest peak in the periodogram obtained for the measurements using the entire 
spectrum is detected at the frequency
0.30\,1/d corresponding to a period of 3.33\,d, while the highest peak at the frequency 0.265\,1/d corresponding to a 
period of 3.77\,d is identified in the periodogram obtained for the measurements 
carried out excluding the \ion{He}{ii} $\lambda$4486 line.
 
\begin{figure}
\centering
\includegraphics[angle=0,width=.4\textwidth]{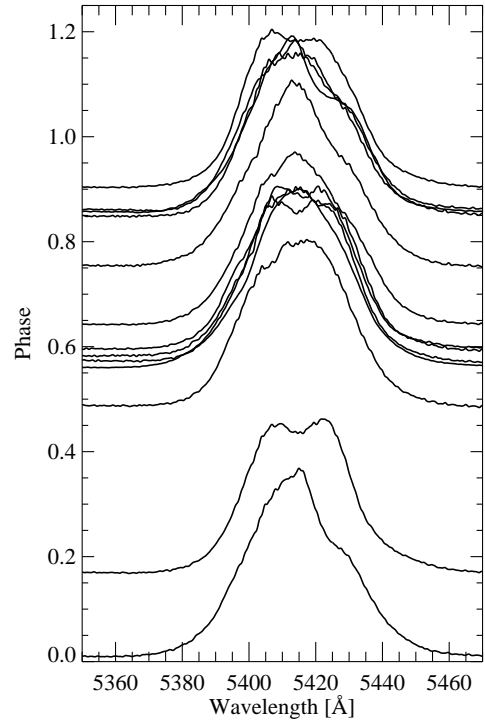}
\caption{
The variability of the \ion{He}{ii} $\lambda$5412 line profile over the rotation cycle.
Phased Stokes~$I$ profiles are presented with the continuum level set to the rotational phase.
For better visibility, all profiles have been scaled by a factor of 0.2.
}
\label{fig:lines}
\end{figure}

In the course of our study, we noted that variations detected in emission 
line profiles probably correlate with the variation of the magnetic field.
As an example, we present in Fig.~\ref{fig:lines} the variability of the shape 
of the \ion{He}{ii} $\lambda$5412 line 
profile, which shows some kind of double-peak structure in the phases close to the magnetic field extrema, i.e.\ close to
the rotation phases 0 and 0.5.
Although our data are rather sparse, it is possible that wind structure variations related to the 
presence of a magnetic field may
lead to the observed changes in the shape of the spectral lines.
However, the line profile 
recorded in 2010 at the phase 0.560 (i.e.\ 77 rotation cycles before the start of our service observations), which 
is close to the time of the maximum of the 
longitudinal magnetic field, does not show this structure. Such instable behaviour of the 
line profiles in the spectra separated in time by several months or years was previously reported in the study of 
\citet{flores2011}.
Extensive sets of optical spectroscopic observations spread over years have been 
obtained in the past by several groups, mainly aimed at studies of the relationship between the level of
continuum flux emerging from the inner stellar wind and the spectroscopic changes. Despite the epoch dependency
of the variations,
\citet{morel1998} found a persistent correlation between changes in the wind line profiles 
and in the continuum emission
forming at the pseudo-photosphere in the wind.

\section{Measurement error validation with simulated data}
\label{sect:syn}

An independent estimate of the error margin of our magnetic field measurements using circular polarization 
measurements of WR\,6 can be obtained from the following test based on simulated data. With the Potsdam 
Wolf-Rayet (PoWR) model atmosphere code \citep{Grafener2002,Hamann2003}, we calculated 
a synthetic spectrum that is similar to the observed spectrum of this star, presented in Fig.~15
in the work of \citet{hamann2006}.

The advantage of using synthetic models is that they are created without noise and therefore one can add 
a predetermined amount of noise according to the goodness of the data. At first, we take a normalized model 
spectrum and reduce the resolution to the FORS\,2 standard 2.4\,\AA{} FWHM and also bin the pixel size according 
to the dispersion of 0.75\,\AA{}. Then, we add Gaussian distributed noise with a specified SNR. We selected a 
SNR similar to the observation in the night of 2010 December 24.
This procedure is done twice so that one model spectrum with noise can be taken as right circular polarized and
the other as left circular polarized. Then the data are analyzed for their Zeeman shift in the same way as the 
real observations.

\begin{figure} 
 \centering
   \includegraphics[angle=270,scale=0.39]{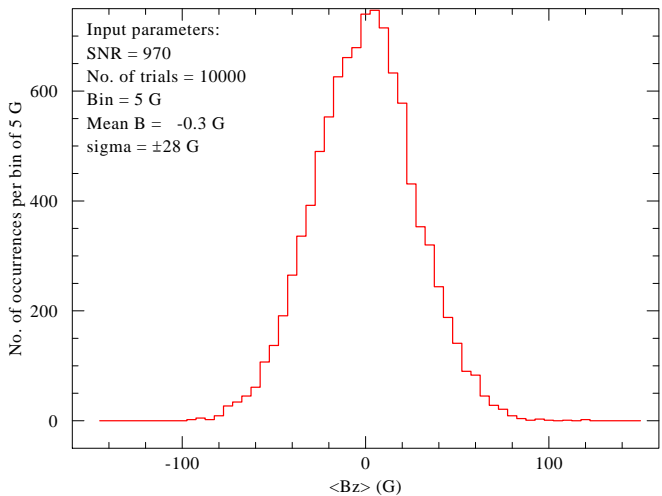}
    \caption{Distribution of $10\,000$ $\left<B_{\rm z}\right>$ measurements on simulated data for WR 6 with 
input field $\left<B_{\rm z}\right> = 0$ and artificial noise corresponding to a SNR of 970 per pixel in 
each channel. The result shows a 1$\sigma$ error of 28 G.}
    \label{fig:error0}
\end{figure}

This has been done with a SNR of 970 and repeated for $n=10\,000$ times in order to get a normal distributed 
value of the null field from which it is possible to calculate the standard deviation. This gives an estimate 
of the expected error for the measurement of the longitudinal magnetic field. The distribution of the 
measured $\left<B_{\rm z}\right>$ values is plotted as a histogram in Fig.~\ref{fig:error0}. They scatter 
around zero with a mean deviation of 28\,G.  

Summarizing, this test was based on the following assumptions: (1) the line spectrum is similar to our 
simulated, normalized spectrum for WR 6; (2) the observed spectra have a S/N of 970 per pixel in the 
ordinary as well as in the extra-ordinary channel; (3) statistical noise is the only source of errors; 
(4) all lines experience the same Zeeman splitting. 
Under these assumptions, the value obtained for $\left<B_{\rm z}\right>$ with FORS2 via our method of
analysis has a theoretical 1$\sigma$ error of $\sim28$ G. Note that this error is independent of the 
magnetic field strength assumed in modeling the simulated Stokes~$V$ spectrum (here $\left<B_{\rm z}\right> = 0$).

\section{Linear polarization}
\label{sect:lin}

\subsection{Linear polarizations of BAT99\,7, WR\,7, WR\,18, and WR\,23}

Apart from the stars WR\,6 and WR\,18, no linear polarization observations are reported in the literature for 
BAT99\,7, WR\,7, and WR\,23.
For WR\,18, we found in the article of \citet{schulte1994} a note that the linear polarization spectrum
of this star is featureless. Previous studies of linear polarization of WR\,6 indicated the presence of a ``line effect'', 
whereby the polarization across strong lines is generally smaller than in neighboring regions of the continuum 
(e.g., \citealt{mclean1979}; 
 \citealt{schulte1990,schulte1991}; \citealt{harries1999}).

\begin{table*}
\caption[]{Continuum linear polarization measurements.
}
\centering
\begin{tabular}{ll r@{$\pm$}l r@{$\pm$}l r@{$\pm$}l r@{$\pm$}l l}
\hline
\hline 
\noalign{\vspace{1mm}}
\multicolumn{1}{c}{Object} &
\multicolumn{1}{c}{MJD} &
\multicolumn{2}{c}{$Q$} &
\multicolumn{2}{c}{$U$} &
\multicolumn{2}{c}{$P$} &
\multicolumn{2}{c}{$\theta$}& 
\multicolumn{1}{c}{Phase} \\
 &
 &
\multicolumn{2}{c}{\%} &
\multicolumn{2}{c}{\%} &
\multicolumn{2}{c}{[\%]} &
\multicolumn{2}{c}{[$^{\circ}$]} &
\\
\noalign{\vspace{1mm}}
\hline
BAT99\,7 & 55554.1999&0.52&0.04 &0.78&0.04& 0.94 & 0.04 & 28.16 & 1.22 &\\
BAT99\,7 & 55555.1293&0.49&0.06&0.75 &0.06 &0.90 & 0.06 & 28.42 & 1.92 &\\
WR\,7    & 55554.2874&1.23&0.03 &$-$1.76&0.03& 2.15 & 0.03 & $-$27.53& 0.40 &\\
WR\,7    & 55555.2473&1.27&0.04 &$-$1.74 &0.04& 2.15 & 0.04 & $-$26.93& 0.53 &\\
WR\,18   & 55555.3267&$-$0.20&0.03 &$-$1.91&0.03& 1.92 & 0.03 & $-$47.99& 0.45 &\\
WR\,23   & 55554.3465&$-$2.66&0.04 &$-$2.99&0.04& 4.00 & 0.04 & $-$65.83& 0.29 &\\
WR\,23   & 55555.3596&$-$2.66&0.05&$-$3.01 &0.05& 4.02 & 0.05 & $-$65.73& 0.36 &\\
WR\,6    & 55555.1672&0.61&0.06 &0.02&0.06& 0.61 & 0.06 & 0.94 & 2.82 & 0.563 \\
WR\,6    & 55845.2830&0.22&0.05 &$-$0.18&0.05& 0.28 & 0.05 & $-$19.64 & 5.04 & 0.599\\
WR\,6    & 55846.2701&$-$0.39&0.09 &$-$0.31&0.09& 0.50 & 0.09 &$-$ 70.76 & 5.18 & 0.861\\
WR\,6    & 55880.1753&$-$0.41&0.08 &$-$0.25&0.08& 0.48 & 0.08 &$-$ 74.31 & 4.77 & 0.864\\
WR\,6    & 55903.3289&$-$0.28&0.08 &0.15&0.08& 0.32 & 0.08 &75.91& 7.22 & 0.012\\
WR\,6    & 55906.1369&0.05&0.07 &$-$0.38&0.07& 0.38 & 0.07 & $-$41.25 & 5.23 & 0.758\\
WR\,6    & 55909.2286&0.32&0.07 &0.02&0.07& 0.32 & 0.07 & 1.79 & 6.25 & 0.578\\
WR\,6    & 55928.0893&0.35&0.05 &$-$0.15&0.05& 0.38 & 0.05 & $-$11.60 & 3.76 & 0.587\\
WR\,6    & 55929.1006&$-$0.40&0.09 &0.32&0.09& 0.51 & 0.09 & 70.67 & 5.03 & 0.855\\
WR\,6    & 55932.0836&0.27&0.05 &$-$0.14&0.05& 0.30 & 0.05 & $-$13.70 & 4.71 & 0.647\\
WR\,6    & 55933.0627&0.04&0.05 &0.48&0.05& 0.48 & 0.05 & 42.62 & 2.94 & 0.907\\
WR\,6    & 55934.07260&0.25&0.06 &0.03&0.06&0.25 & 0.06 & 3.42 & 6.83 & 0.175\\
WR\,6    & 55935.2602&0.28&0.07 &0.15&0.07& 0.32 & 0.07 & 14.09 & 6.31 & 0.491\\
\noalign{\vspace{1mm}}
\hline         
\label{tab:linpol}
\end{tabular}  
\end{table*} 

In the upper part of Table~\ref{tab:linpol}, we present the results of the measurements of linear 
polarization in the wavelength range between
4000 and 6100\,\AA{} for BAT99\,7, WR\,7, WR\,18, and WR\,23.
For all these stars, the polarization spectra are featureless, i.e.\ no evidence of a line effect was 
found.
The WR star BAT99\,7 is a WN4b type star in the LMC, characterised by very round line 
profiles. As mentioned in the Introduction, such line profiles are believed to be 
explained by very rapid rotation that would lead to a deformation of the star.
A large deviation from a 
spherical shape is expected to produce continuum polarization through scattering effects and only the line 
emission would be less scattered as it originates farther out in the wind. 
Unfortunately, no line effect is detectable in our linear polarization measurements. 

A few years ago, \citet{chene2011} undertook a search for spectral variability in a large sample
of WR stars, among them the stars WR\,7, WR\,18, and WR\,23. While WR\,7 and WR\,18 showed no 
detectable spectral variability, 
the star WR\,23 was classified as a star showing small-scale profile variability.  

\begin{figure}
\centering
\includegraphics[angle=0,width=.45\textwidth]{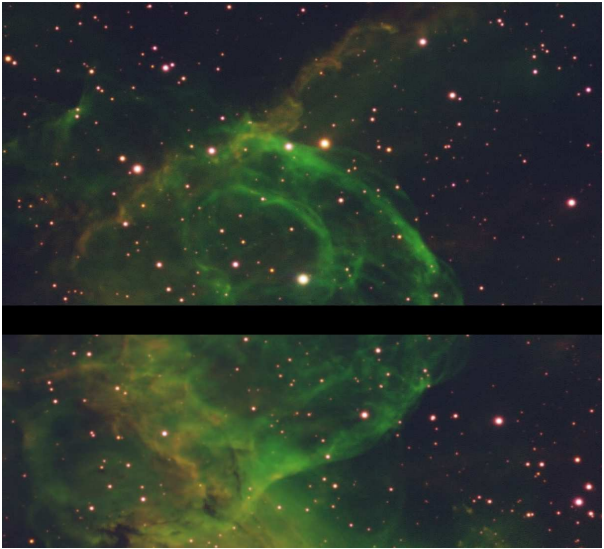}
\caption{Combined FORS\,2 image of the nebula NGC\,2359 around the star WR\,7. The detector is an E2V mosaic of 
two 2k$\times$4k CCDs with a field of view of 6.8$\arcmin$$\times$6.8$\arcmin$.
The black strip close to the middle of the image is due to a gap between the two chips. 
}
\label{fig:wr7}
\end{figure}

In the observed WR stars, all measured polarization is the sum of interstellar polarization (ISP) and intrinsic 
polarization. Usually, for stars without a line effect, the polarization
properties of nearby field stars are used to estimate the ISP and its position angle.
Importantly, 
narrow-band optical surveys of the environments of WR stars indicate the presence of nebulae around WR\,7, WR\,18, and WR\,23
\citep{Marston1997}.
As an example, we present in Fig.~\ref{fig:wr7} an image of the nebula NGC\,2359 
around the star WR\,7, obtained during the target acquisition with FORS\,2. 
This image was obtained by 
combination of the broad-band $b_{\rm high}$ filter with the narrow-band
\ion{S}{ii} and  H$\alpha$ filters.
An interstellar gas shell was also observed in the direction of the WR star WR\,6,
but presently it is not completely clear whether there is a physical connection between this object and the shell
\citep{fesen1994}. 
Due to the presence of nebulae around our WR targets, we expect that field stars in their vicinity  
do not probe the same interstellar material and do not display the
same ISP. Therefore, no attempt has been made to correct the data for 
interstellar polarization and the values presented in Table~\ref{tab:linpol} cannot be considered as intrinsic 
to these stars. On the other hand, three WR stars were observed on two different nights, so that our analysis 
should allow us to detect the presence of variable polarization that is intrinsic to the star. 
Taking into account the measurement accuracy, no spectropolarimetric variability is 
detected between the two nights for any star.  
The star WR\,23 shows the largest amount of polarization of about 4\%, while the smallest polarization 
was measured for BAT99\,7.

\subsection{Interstellar and intrinsic linear polarization of WR\,6}
\label{sect:linpolwr6}

\begin{figure}
\centering
\includegraphics[angle=0,width=.45\textwidth]{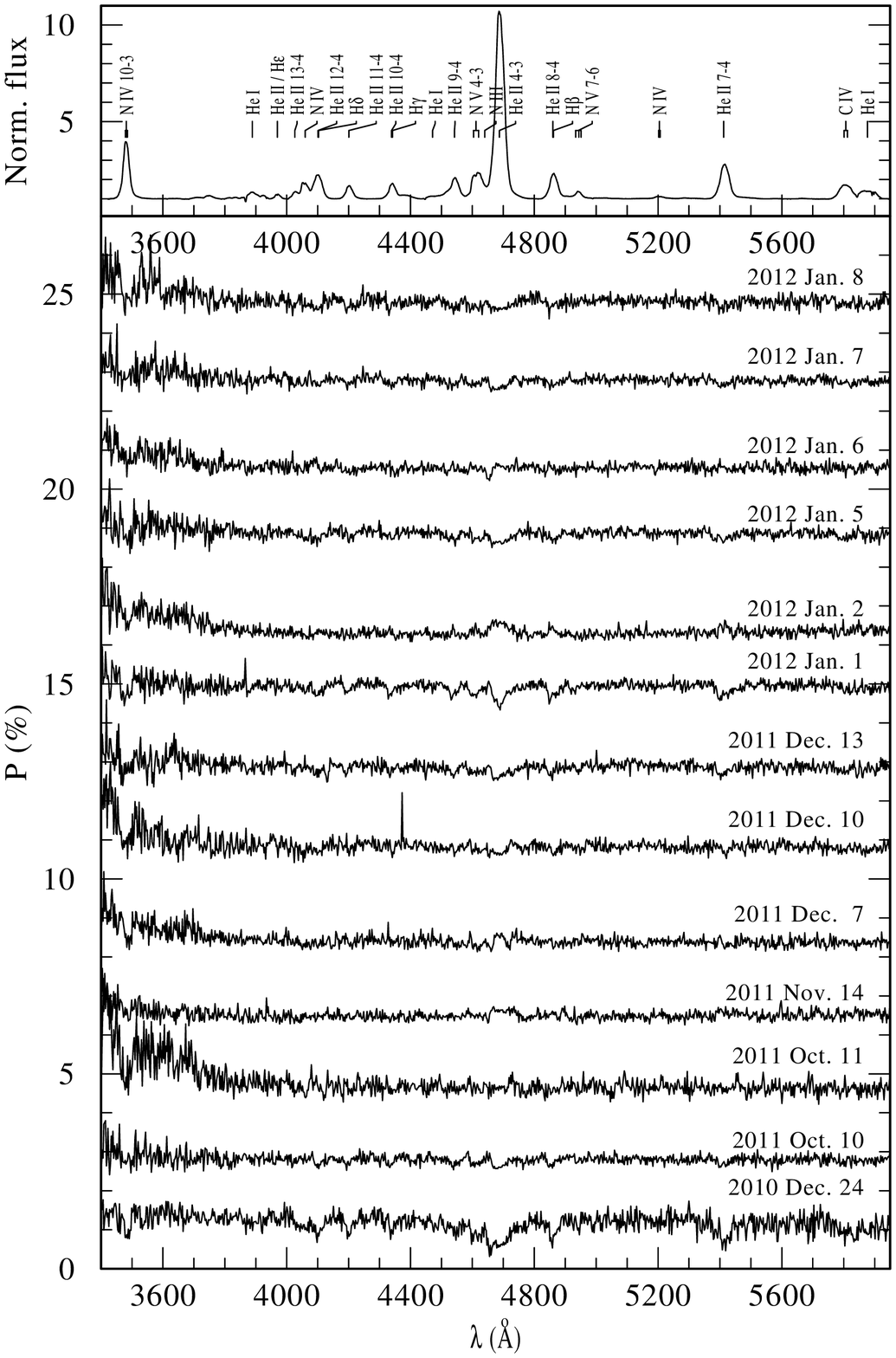}
\caption{Linear polarization spectra of WR\,6 not corrected 
for the interstellar polarization. Each spectrum is shifted in vertical direction by 2\% for 
a better visibility.
}
\label{fig:linwr6}
\end{figure}

\begin{table}
\caption[Spectral regions]{
Spectral regions that were used to measure the mean value of the line and continuum polarization.
}
\centering
\begin{tabular}{lcc}  
\hline
\hline 
\noalign{\vspace{1mm}}
\multicolumn{1}{c}{Line} &
\multicolumn{1}{c}{line region} &
\multicolumn{1}{c}{continuum region} \\
 &
\multicolumn{1}{c}{[\AA{}]} &
\multicolumn{1}{c}{[\AA{}]} \\
\noalign{\vspace{1mm}}
\hline
\noalign{\vspace{1mm}}
\ion{He}{ii} 4201 & 4196 - 4208 & 4250 - 4270 \\
\ion{He}{ii} 4339 & 4334 - 4344 & 4270 - 4290 \\
\ion{He}{ii} 4542 & 4537 - 4547 & 4430 - 4440 \\
\ion{He}{ii} 4859 & 4855 - 4869 & 5080 - 5120 \\
\ion{He}{ii} 5412 & 5407 - 5421 & 5560 - 5730 \\
\hline
\end{tabular}
\label{tab:lin}
\end{table}

\begin{figure}  
\centering
   \includegraphics[angle=270,width=0.45\textwidth]{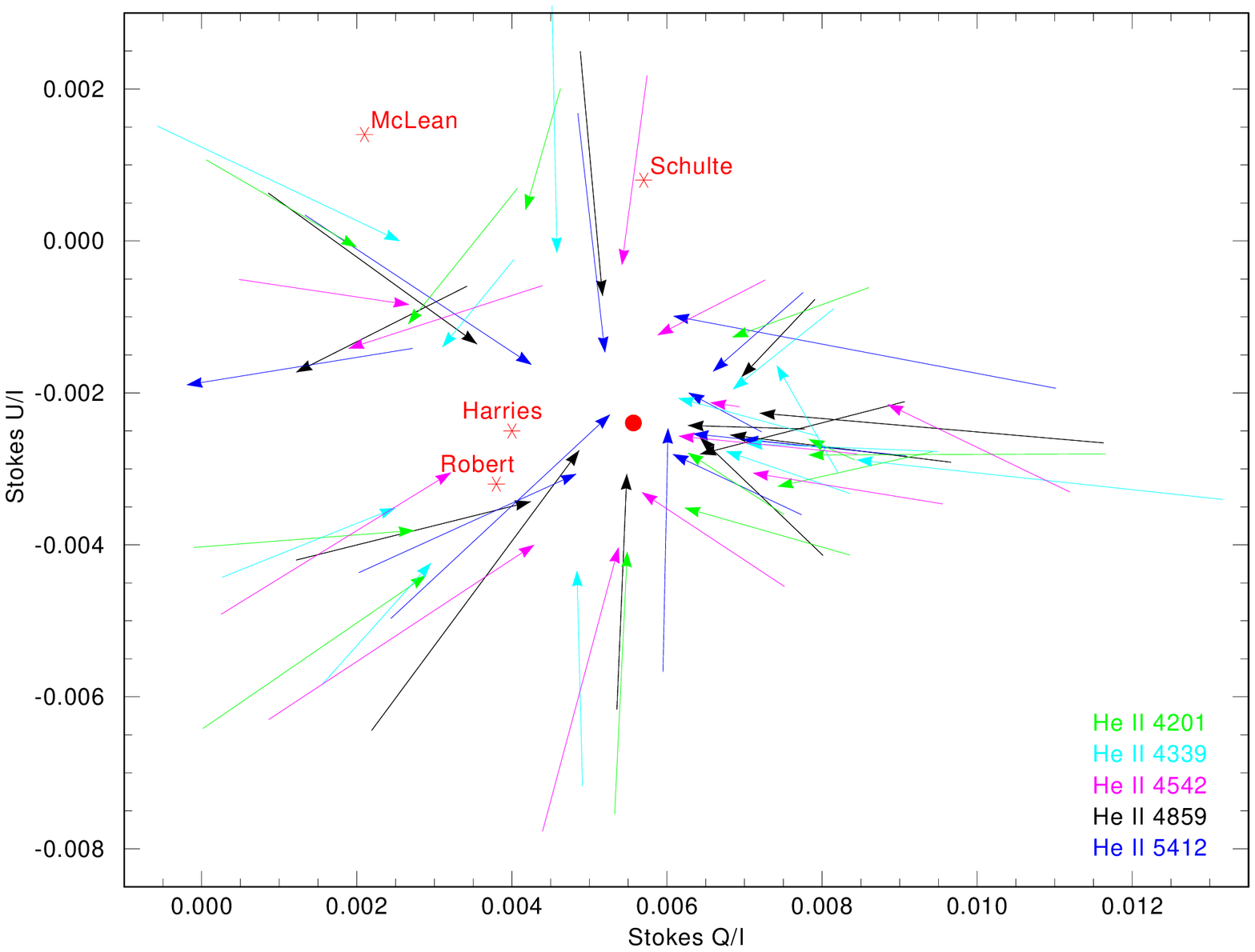}
   \caption[Interstellar polarization of WR\,6]{
Interstellar polarization of WR\,6 determined from the 
polarization vectors is marked by a filled circle.
Previous measurements were carried out by \citet{schulte1991}, 
\citet{mclean1979}, \citet{robert1992}, and \citet{harries1999},
identified by the individual stars.
Vectors are shown for the lines
\ion{He}{ii} 4201 (light green),
\ion{He}{ii} 4339 (light blue), 
\ion{He}{ii} 4542 (magenta),
\ion{He}{ii} 4859 (black), and
\ion{He}{ii} 5412 (dark blue).
}
   \label{fig:lin_isp_wr6}
\end{figure}

 Since all measured polarization is the sum of interstellar polarization and intrinsic 
polarization (whereby the line polarization is always a fraction of the continuum polarization),
to determine the interstellar polarization in WR\,6, we followed the procedure described in the work of \citet{harries1999}. 
In the stellar spectra displaying the line effect, it is possible to obtain a good measure of the 
interstellar polarization. In order to estimate it, one needs to determine the polarization of the 
line and the continuum. In Fig.~\ref{fig:linwr6}, we present as an example linear polarization spectra not corrected 
for interstellar polarization. The continuum polarization of WR\,6 was calculated 
using the vector sum of the Stokes parameters over the entire spectrum, excluding
regions containing obvious emission lines. 
The respective wavelength regions used are presented in Table~\ref{tab:lin}. 
This results in a vector pointing from the continuum polarization to the line polarization and finally 
to the direction of the interstellar polarization
(see Fig.~\ref{fig:lin_isp_wr6}).
Then, we assumed for 
each observation a line that goes through the line polarization, the continuum polarization and an 
initial interstellar polarization (ISP). Through a least-squares minimization for each line and a minimization of $\sum \chi^2$ 
via a downhill simplex technique, a best estimate for ($q_i$, $u_i$) could be determined. We obtained 
$q_i=0.56\pm0.08$ and $u_i=-0.24\pm0.08$, which gives $P=0.61\pm0.08\%$ and $\theta=-11.60\pm3.76^{\circ}$. The interstellar 
polarization is wavelength dependent and can be modelled by the empirical Serkowski law \citep{serkowski1973}:

\begin{equation}
 P(\lambda)=P_{\textrm{max}} \exp\left[-k \ln^2\left(\frac{\lambda_{\textrm{max}}}{\lambda}\right)\right]
\end{equation}

\noindent
where $P_{\textrm{max}}$ is the maximum polarization occurring at a wavelength 
$\lambda_{\textrm{max}}$ and $k\sim1.15$ is a shape parameter.  

The results of the measurements of linear polarization in the spectra of 
WR\,6 are presented in the lower part of Table~\ref{tab:linpol}, together with the corresponding phases.
The linear polarization of the star WR\,6 was already presented in numerous studies (e.g.,
\citealt{schulte1991}; \citealt{mclean1979}; 
\citealt{robert1992}; \citealt{harries1999}),
and our measurements show close agreement with previous results: we find that the range of the intrinsic 
continuum polarization is approximately 0.2--0.6\%, while the
position angle of the polarization is strongly variable.
We noticed at the rotation phase 0.855 
an apparent emission resulting from the alignment of the continuum to the line polarization and 
the ISP vector (e.g.\ \citealt{harries1999}).
We also noticed that at the rotational phase 0.012 the vector from the continuum to the line polarization 
significantly differs from all other measurements.

\subsection{Modeling a corotating interaction region in WR\,6}

As mentioned above, WR\,6 displays variability that cycles on a period of 3.77\,d.
This period has generally been associated with rotation, and the star
is believed to host a CIR as suggested
by UV, optical studies, and X-rays (e.g.,\ \citealt{louis1995}; 
\citealt{morel1997}; \citealt{ignace2013}).
The stable period of the star indicates the presence of a clock in the system.  However, the variability can exhibit 
complex behavior, including rapid phase changes (e.g., \citealt{ant1994}).  If the period observed in WR\,6 is associated 
with rotation, then the observed variability is tied to an organized structure threading the wind, such 
as CIRs.
In terms of the period, the linear
polarization observations from this study span nearly 100 cycles, and
variability in WR~6 is known to exhibit phase drift. However, much of
the data were obtained within two contiguous cycles.  We have attempted
a rough fit to this subset of the linearly polarized data using a CIR
model as described in \citet{ignace2015}.

\citet{ignace2015} presented a kinematic description of a CIR in terms of
a spiral arm morphology in shape, and in terms of a density contrast
(either enhancement or deficit) relative to the otherwise unperturbed
spherical wind.  This is clearly a gross simplification as compared to
more detailed numerical simulations of CIR phenomena from massive star
winds (e.g., \citealt{cranmer1996}; \citealt{dessart2004}). Using this approach,
\citeauthor{ignace2015} employ the optically thin scattering results of 
\citet{brown1977} for axisymmetric structures. The spiral morphology is not
axisymmetric; however, the structure is conceived in terms of a series
of differential pancake-shaped blobs, each one being slightly shifted
in azimuth from the preceding, so that each segment of the spiral is
axisymmetric. Using an equation of motion for the spiral pattern, and
formulating a superposition for the net polarization properties from the
ensemble of spiral segment pieces, \citeauthor{ignace2015} were able to evaluate
a semi-analytic result for the polarization with rotational phase.
Moreover, an approach for combining multiple CIRs originating from
arbitrary latitudes was presented.

Here only a heuristic application of the model is applied to the dataset.
The model allows for a number of free parameters, but in the application
to WR~6, the star and wind parameters are treated as fixed. The wind
is optically thick to electron scattering, and the Ignace et al.\ model
applies only for thin scattering. For this situation a core-halo model is
adopted, whereby the polarization is evaluated only where the electron
optical depth $\tau_{\rm e} < 1$, assuming that light from regions with
$\tau_{\rm e} > 1$ will be completely depolarized.

The free parameters that remain in the model include: viewing inclination
$i$ (relative to the spin axis of the star), latitude $\vartheta$ at
which a CIR originates, opening angle $\beta$ of the CIR structure,
the rotational phase offset $\varphi_0$ between the model and the
observations, an orientation phase $\psi_0$ for how the projected spin
axis is projected on the sky relative to the observational system
for evaluating Stokes $Q$ and $U$, and finally a density contrast
factor $\eta$. Note first that $\eta$ is defined so that $\eta=0$ is
a spherical wind. Second, the factors $\beta$ and $\eta$ are somewhat
degenerate when the opening angle of the CIR is not overly large:
both affect the amplitude of the polarization but not the shape of its
variation with rotational phase.

\begin{figure}
\centering
\includegraphics[angle=0,width=.45\textwidth]{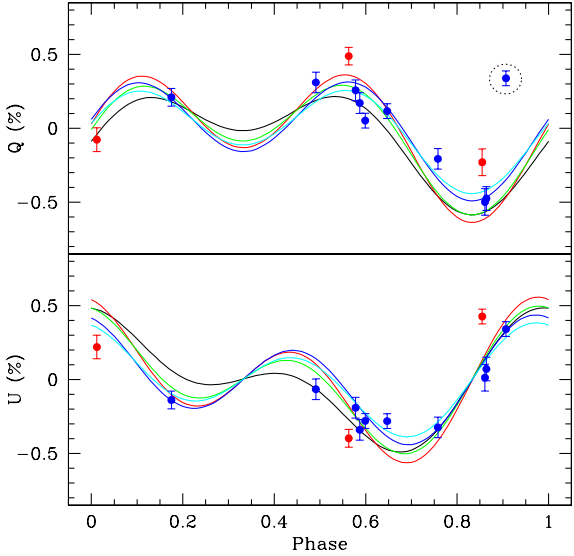}
\caption{
The two panels present linear polarization $Q$ (upper panel) and $U$ (lower panel) in  percent
plotted against rotational phase.  The points present the data while the curves
correspond to the models for viewing inclinations of $i =150^\circ - 165^\circ$.
}
\label{fig:modell}
\end{figure}

Although the model allows for multiple CIRs, our objective was to
determine whether a reasonable match between the data and the model
could be achieved with just one CIR structure.  The results of our
trial-and-error fitting are shown in Fig.~\ref{fig:modell}.  The two
panels are for $Q$ (upper panel) and $U$ (lower panel) as percent
polarization plotted against rotational phase.

Note that formal fits between the data and the model are not made (i.e.,
there is no $\chi^2$ evaluation).  Also, the three red points in Fig.~\ref{fig:modell} were not used
in guiding the selection of the model parameters.  Of those three points,
one is from a much older observation at the phase 0.563, and thus from a much earlier epoch than
the blue points.  The other two points that were not considered are the point at rotation phase 0.855,
where an apparent emission was resulting from the alignment of the continuum to the line polarization and 
the ISP vector, and the point at rotation phase 0.012, where the vector from the continuum to the line polarization 
significantly differed from all other measurements (see Sect.~\ref{sect:linpolwr6}).
However, as is shown in Fig.~\ref{fig:modell},
all three points do roughly follow the trend of the model curves.

There is one notable discrepant point in the panel for $Q$,
located near phase 0.9 (surrounded by a dotted circle).
No model with a single arm can simultaneously account for that point
along with the other measurements. The discrepant point may suggest more
arms, or that a CIR (or CIRs) is not the only structure existing in
the flow. We should expect a stochastic component to the structure.
The CIR might be disrupted at times, or perhaps an additional wind structure
can sometimes dominate the polarization.

In the selecting the model curves plotted in Figure~\ref{fig:modell},
we first tried a single CIR located in the equatorial plane. Varying the
remaining parameters, no satisfactory matches were achievable. Next we
varied both the latitude of the CIR and the viewing inclination of the
star. A suite of model curves for viewing inclinations of $i =150^\circ - 165^\circ$ 
are displayed in Fig.~\ref{fig:modell}, which are seen
to broadly reproduce the pattern for the data (with the exception of
one point in $Q$ near phase of 0.9). Relative to inclinations below
$90^\circ$, an inclination exceeding $90^\circ$ serves to flip the
sense of rotation as projected onto the sky. For this range a fairly
similar set of model curves resulted for CIRs at latitudes of $40^\circ$
and $45^\circ$ as well as $135^\circ$ and $140^\circ$. When $i$ and
$\vartheta$ are in opposite hemispheres, the spiral does sweep across the
front of the star for the observer's sightline, but WR~6 is known to show
regular DAC behavior (e.g.\ \citealt{louis1995}).  Figure~\ref{fig:modell}
thus shows the models for $\vartheta= 135^\circ$ and $140^\circ$.

This is the first time that a model for polarization from a CIR has been
applied to data.  Given that the model adopts a core-halo approach,
plus the complexity of the wind structure, the overall ability of
the model to match the majority of the data suggests that a CIR is
plausibly the source, or a significant contributor, to the variable
linear polarimetry of WR~6. It is not our goal to claim a fit of high
significance, but simply to test the ability of the model for achieving a
rough match to the observed trend with (a) a single CIR and (b) a limited
number of free parameters.  Given the simplistic nature of the model,
it is gratifying to find that a modestly good fit to the majority of
the measures can be achieved.

\section{Discussion}
\label{sect:disc}

In this work, we address Wolf-Rayet type
stars, where the magnetic fields are especially hard to detect because of the
wind-broadening of their spectral lines. Moreover, all photospheric lines are absent and 
the magnetic field is measured on emission lines formed
in the strong wind.

It is the first time that the low dispersion spectrograph FORS\,2 in spectropolarimetric 
mode mounted at one of the biggest telescopes in the world is used for spectropolarimetry of WR stars. 
WR stars descend from initial
masses above 20\,$M_\odot$ and during their fast evolution, they already
lost a significant mass fraction via their powerful stellar winds.
Depending on their exact mass, rotation, magnetism and mass-loss, in
a few $10^5$\,yr, these stars will become either neutron stars or black
holes (e.g.\ \citealt{heger2003}). A small fraction of the
WR star progenitors have strong magnetic fields on the main sequence --
the magnetic O-type stars. The observed temporal variations of the longitudinal
magnetic fields in these stars are usually compatible with dipole fields
inclined to the rotation axis, implying a simple global structure.
Obviously, the strength of the average
longitudinal magnetic field in WR\,6, varying with an amplitude of order only 100\,G over the rotation cycle,
is significantly lower than the field strength detected in several magnetic O-type stars.
This is perhaps not surprising, given the fact that 
the magnetic field is diagnosed in the line-forming regions, which fall fairly far out in
a Wolf-Rayet wind, and not 
at the stellar surface. 
On the other hand, a Wolf-Rayet star like WR\,6 is significantly contracted relative to its O-star
progenitor, so the evolution of the field between these states requires further investigation.
It would be worthwhile to obtain in the future more dense rotation phase
coverage by spectropolarimetric observations to further strengthen the evidence for the magnetic 
nature of this star.
We should keep in mind that
the method we have applied involves substantial systematic uncertainties, so
only gives a rough result for the field strength where lines are formed, and that different lines form over 
different zones of the wind, sampling different field strengths.

Magnetic fields for the  other WR stars in our sample present at a significance 
level of about 2.5$\sigma$ and below.
Since up to now no long-time spectropolarimetric monitoring was carried out for them and we cannot neglect
the possibility of unfavorable sampling of the magnetic field configuration, the non-detection of the magnetic 
field does not mean that no magnetic field is present in their atmospheres. Clearly, attaining higher SNR
using FORS\,2 in the future will help to set more stringent limits to their magnetic field strengths.

For the first time, we present linear polarization measurements for BAT99\,7, WR\,7, WR\,18, and WR\,23. 
We do not find any evidence of a line effect in these stars.
Previous linear polarization studies showed that most
WR stars exhibit variable continuum polarization. However, no spectropolarimetric variability is 
detected for BAT99\,7, WR\,7, and WR\,23 over two nights.
From the monitoring of the linear polarization of WR\,6 over the rotation cycle, we confirm that the detected 
continuum polarization and the polarization position angle are not constant with time. 


\section*{Acknowledgments}
The authors thank the referee for useful comments and helpful suggestions that improved this manuscript.
We also thank Yuri Beletsky from Las Campanas Observatory for his help in the preparation of the image of the nebula NGC\,2359.


\label{lastpage}

\end{document}